\documentclass[twocolumn,showpacs,preprintnumbers,amsmath,amssymb,aps,prb,superscriptaddress]{revtex4-1}

\pdfoutput=1
\usepackage[pdftex]{graphicx}
\usepackage[english]{babel}
\usepackage{soul}
\usepackage{cleveref}
\usepackage{bbold}
\usepackage{mathtools}
\usepackage{multirow}
\usepackage{colortbl}
\definecolor{kugray5}{RGB}{224,224,224}
\usepackage[latin1]{inputenc}
\usepackage{diagbox}

\usepackage{dcolumn}
\usepackage{bm}

\usepackage{color}
\usepackage{amstext}
\usepackage{braket}
\usepackage{mathdots}
\usepackage{tikz}
\usepackage{physics}
\usepackage{enumitem}

\usetikzlibrary{matrix,decorations.pathreplacing}

\begin{document}


\title{Analytical solution of open crystalline linear 1D tight-binding models}

\author{A. M. Marques}
\email{anselmomagalhaes@ua.pt}
\affiliation{Department of Physics $\&$ I3N, University of Aveiro, 3810-193 Aveiro, Portugal}
\author{R. G. Dias}
\affiliation{Department of Physics $\&$ I3N, University of Aveiro, 3810-193 Aveiro, Portugal}

\date{\today}


\begin{abstract}
A method for finding the exact analytical solutions for the bulk and edge energy levels and corresponding eigenstates for all commensurate Aubry-Andr\'e/Harper single-particle models under open boundary conditions is presented here, both for integer and non-integer number of unit cells.
The solutions are ultimately found to be dependent on the behavior of phase factors whose compact formulas, provided here, make this method simple to implement computationally.
The derivation employs the properties of the Hamiltonians of these models, all of which can be written as Hermitian block-tridiagonal Toeplitz matrices. 
The concept of energy spectrum is generalized to incorporate both bulk and edge bands, where the latter are a function of a complex momentum.
The method is then extended to solve the case where one of these chains is coupled at one end to an arbitrary cluster/impurity.
Future developments based on these results are discussed. 
\end{abstract}


\maketitle

\section{Introduction}
\label{sec:intro}

Tight-binding (TB) models have become ubiquitous due to their success in describing a wide array of different physical systems, ranging from condensed matter \cite{Goringe1997} to photonic lattices \cite{Garanovich2012} and ultracold atoms in optical lattices \cite{Lewenstein2012}, etc. 
For $d$-dimensional ($d$D) crystalline models under periodic boundary conditions (PBC), finding the energy spectrum associated to the Bloch eigenstates is a straightforward task.
Complications arise, on the other hand, when open boundary conditions (OBC) are considered.
In this case, the broken translational symmetry at the edges prevents a direct calculation of the energy bands.
Furthermore, edge-localized states, possibly corresponding to symmetry-protected topological (SPT) states, may appear under OBC \cite{Hasan2010,Qi2011}.
Already at the level of open 1D TB models, general formulas for the analytical determination of both the bulk and edge states are hard to find, with exact solutions not going beyond the Su-Schrieffer-Heeger (SSH) model \cite{Delplace2011}, a textbook example of a 1D topological insulator \cite{Asboth2016}.
However, recent progress was made regarding more general crystalline 1D models (or higher-dimensional models with OBC along a single direction) \cite{Kunst2017,Alase2017,Duncan2018,Kunst2019}.
Here, we detail on how to find the analytical solutions of an entire family of 1D linear models under OBC, namely the extensively studied family of (commensurate) Aubry-Andr\'e/Harper (AAH) models \cite{Lahini2009,Kraus2012,Ganeshan2013,Lang2014,Shen2014,Schreiber2015,Ke2016,Zeng2016,Cao2017,Zhao2017,Malla2018,Das2019}, and accordingly general expressions for finding all the eigenstates and corresponding energy spectra are provided here in an easily algorithmizable format.

\begin{figure}[h]
	\begin{center}
		\includegraphics[width=0.47 \textwidth]{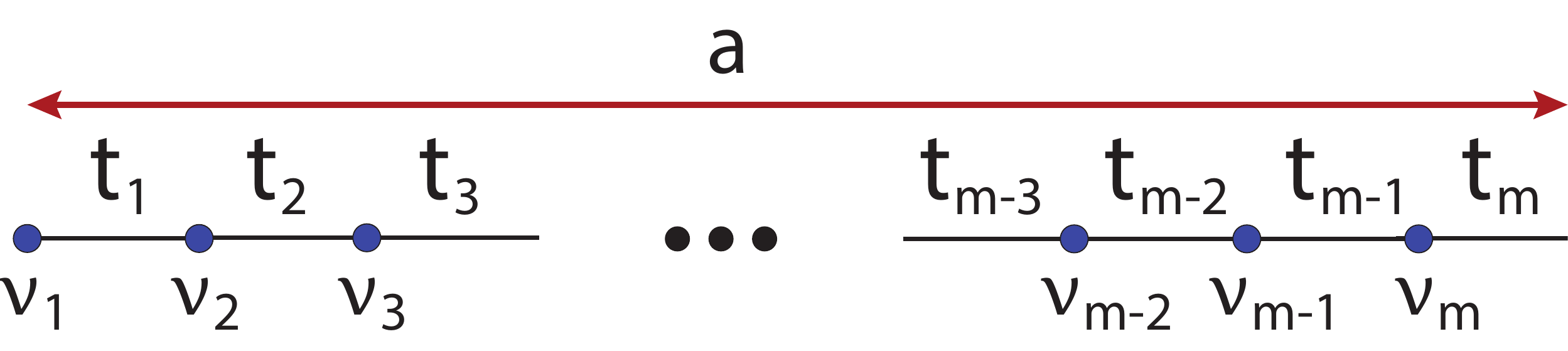}
	\end{center}
	\caption{Unit cell of the ISSH$_m$ model, where $\{t_i\}$ are the hopping parameters, $t_m$ is the intercell hopping parameter, $\{v_i\}$ are the on-site potentials and $a$ is the lattice spacing.
		The SSH$_m$ model is obtained by setting all $\{v_i\}$ in the ISSH$_m$ model to zero.}
	\label{fig:ionicsshm}
\end{figure}

Central to the studies carried out below is that the Hamiltonian of an open 1D linear model with arbitrary on-site potentials $v_i$ and hopping constants $t_i$ within the unit cell, labeled ionic SSH$_m$ (ISSH$_m$), where $m$ is the size of the unit cell (see Fig.~\ref{fig:ionicsshm}), is a periodic Hermitian block-tridiagonal matrix \cite{Rozsa1992}, also called a tridiagonal Hermitian $m$-Toeplitz matrix \cite{Alvarez2005,Fonseca2007,Alvarez2012,Hariprasad2015,Sahin2018}.
Each ISSH$_m$ model corresponds to an AAH model with specific periodic modulations on $t_i$ and $v_i$, which can be different in general, labeled here as commensurate AAH model.
Banchi and Vaia \cite{Banchi2013} showed that the characteristic equation of a model with the same hopping parameter across the chain can be expressed in terms of Chebyshev polynomials of the second kind and, by introducing then edge perturbations \cite{Fonseca2015,Veerman2018} to the system, the authors were able to find exact formulas for the phase shifts these induce on the eigenstates (which were left implicit in a similar study \cite{Eliashvili2014}).
This technique has been proven very powerful in the development of minimal engineering schemes widely adopted in the context of optimizing quantum \cite{Banchi2011,Apollaro2012,Banchi2013b,Francica2016} and classical \cite{Vaia2018} state transfer.
The extension of the method to include midchain impurities \cite{Compagno2015}, whose strength controls the transmission ratio of an incoming wave, can be applied in the generation of NOON states \cite{Compagno2017}.
Here, we show how the method described in [\onlinecite{Banchi2013}] can be extended for general ISSH$_m$ models, selecting some particular cases as pedagogical examples to illustrate the relevant new features.
Even though there is some unavoidable complexity to its rigorous derivation, it is important to highlight that this method ultimately relies on a very simple calculation of phase factors with compact analytical formulas. 
The reader interested in its immediate application can skip directly to Section~\ref{sec:genmet}.
We point out that Eliashvili \textit{et al.} \cite{Eliashvili2017}, by following a different yet analogous approach to the one outlined here, based on the results of [\onlinecite{Beckermann1995}], have already successfully solved the particular cases of the open SSH ($\equiv$ SSH$_2$) and SSH$_4$ models.

The rest of the paper is organized as follows.
In Section~\ref{sec:ssh4}, we familiarize the reader with the method of finding the exact analytical solutions of 1D linear models under OBC by deriving it step-by-step for an illustrative example, namely the SSH$_4$ model.
In Section~\ref{sec:issh}, the method is extended to incorporate models with periodic modulations also on the on-site potentials, with explicit formulas for the ISSH and ISSH$_5$ models presented there.
In Section~\ref{sec:genmet}, general expressions of the method for an arbitrary ISSH$_m$ model are provided in a summarized version.
In Section~\ref{sec:nonintegerucells}, the method is generalized to include ISSH$_m$ chains with OBC and non-integer number of unit cells, that is, with extra sites of an incomplete unit cell added to one of the edges.
In Section~\ref{sec:cluster}, we study edge deformations in the form of arbitrary clusters coupled to an edge site of the ISSH$_m$ model, and determine the momentum shift the deformation induces on the solutions.
Finally, in Section~\ref{sec:conclusions} we conclude and point to possible future developments on the subject.

\section{SSH$_4$ model}
\label{sec:ssh4}
The method for finding the analytical solutions to the eigenvalues and eigenvectors of a crystalline 1D model with a tridiagonal Hamiltonian and open boundaries can be best understood with a hands-on approach. As such, before we generalize the method we start by demonstrating how it is applied to solve the concrete example of the SSH$_4$ model, whose real-space Hamiltonian, under OBC, is written as an Hermitian periodic tridiagonal matrix \cite{Rozsa1992},
\setcounter{MaxMatrixCols}{20}
\begin{equation}
H=-\begin{pmatrix}
0&t_3&&&&&&&&
\\
t_3&0&t_2&
\\
&t_2&0&t_1&
\\
&&t_1&0&t_4
\\
&&&t_4&0&
\\
&&&&\ddots&\ddots&\ddots
\\
&&&&&&0&t_4
\\
&&&&&&t_4&0&t_3
\\
&&&&&&&t_3&0&t_2
\\
&&&&&&&&t_2&0&t_1
\\
&&&&&&&&&t_1&0
\end{pmatrix},
\label{eq:hamiltssh4}
\end{equation}
where
\begin{equation}
U=-\begin{pmatrix}
0&t_3&&
\\
t_3&0&t_2
\\
&t_2&0&t_1
\\
&&t_1&0
\end{pmatrix}
\label{eq:ucellssh4}
\end{equation}
is the periodic unit cell block that is repeated $N$ times, $t_4$ is the intercell hopping and the basis follows the order $\{\ket{4N},\ket{4N-1},\dots,\ket{2},\ket{1}\}$, with $\ket{j}$ the $j^{th}$-site of the chain. 

\begin{figure}[h]
	\begin{center}
		\includegraphics[width=0.47 \textwidth]{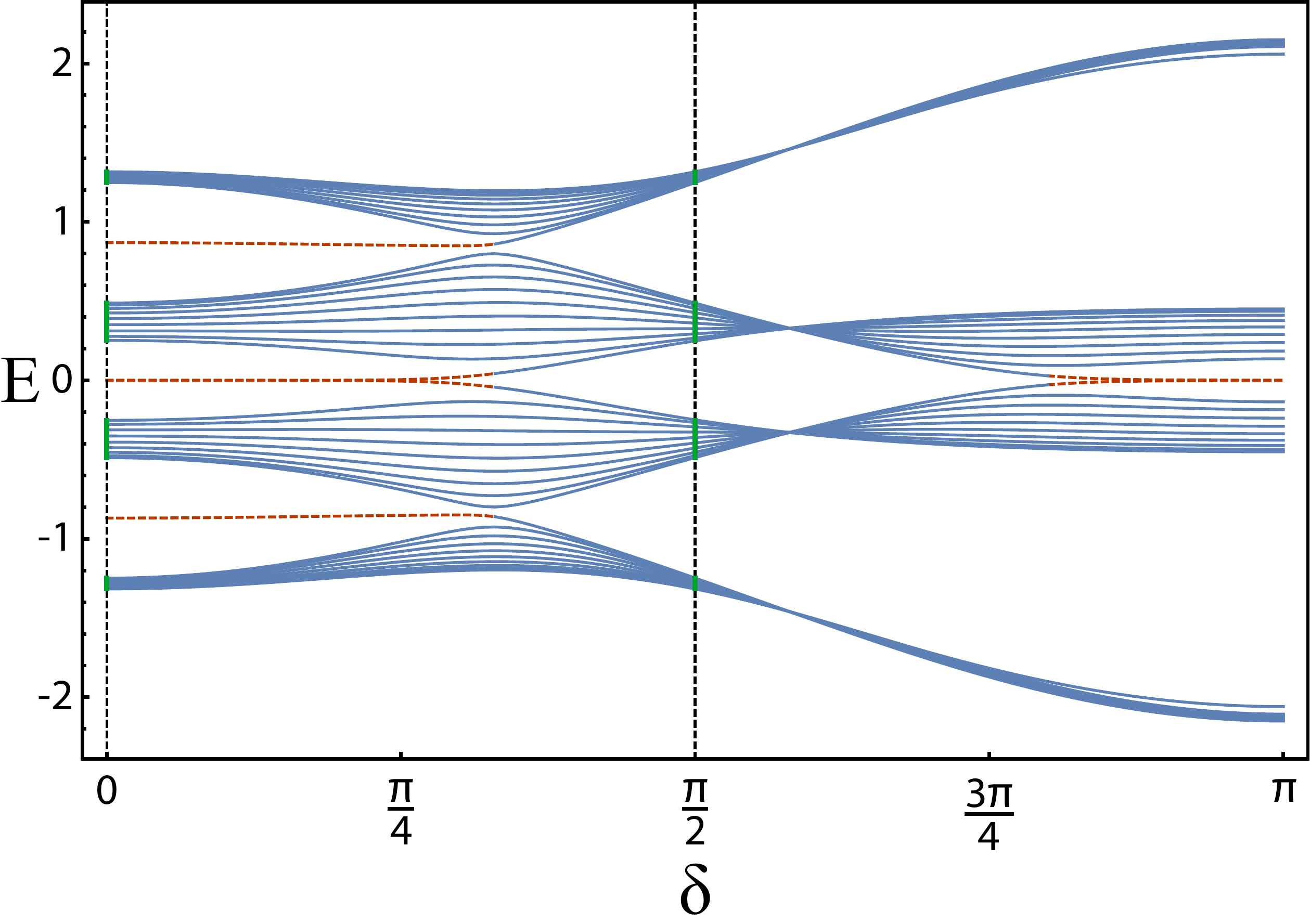}
	\end{center}
	\caption{Energy spectrum in units of $t$ of an open SSH$_4$ chain with $N=10$ unit cells as a function of $\delta$.
		Symmetric $\delta$ have the same spectrum.
		Solid blue curves represent bulk states and the dashed red curves represent in-gap edge states (degenerate for zero energy). At $\delta=0,\frac{\pi}{2}$ the energy spectra (vertical dashed lines) are the same, apart from four in-gap states for $\delta=0$ that become bulk states for $\delta=\frac{\pi}{2}$.}
	\label{fig:espectrumssh4vsdelta}
\end{figure}
For convenience, we introduce a dependence of the hopping parameters of the general SSH$_m$ model on a synthetic momentum $\delta$,
\begin{equation}
t_j(\delta)/t=1-\frac{0.8}{m-1}\Big[(1-\cos \delta)(j-1)+\cos \delta(m-j)\Big],
\end{equation}
where we set $t=1$ as the energy unit and $j=1,2,\dots,m$.
When $\delta=0,\frac{\pi}{2}$, the $t_j(\delta)$ are uniformly spaced between 0.2 and 1, but with opposite progressions.
For example, in the SSH$_4$ model we get
\begin{eqnarray}
\delta=0 &\to& (t_1,t_2,t_3,t_4)=(0.2,0.4\bar{6},0.7\bar{3},1),
\label{eq:deltazero}
\\
\delta=\frac{\pi}{2} &\to& (t_1,t_2,t_3,t_4)=(1,0.7\bar{3},0.4\bar{6},0.2),
\label{eq:deltapiovertwo}
\end{eqnarray}
where the overbars indicate repeating decimals.
The energy spectrum of an open SSH$_4$ model as a function of $\delta$ is given in Fig.~\ref{fig:espectrumssh4vsdelta}, where it can be seen that $\delta=0$ and $\delta=\frac{\pi}{2}$ (the relevant cases from hereafter) have the same spectrum, apart from four states which change from in-gap to bulk states above the gap closing point.

The characteristic polynomial of the whole system is defined as
\begin{equation}
\chi_{_{1:4N}}(\lambda)=\det(\lambda-H),
\label{eq:chissh4general}
\end{equation}
which can be expanded in two ways: i) a top down expansion, growing from $\chi_{_{1:1}}$ (a chain with a single site at position $4N$) to $\chi_{_{1:4N}}$ (the complete chain with all $4N$ sites), or ii) a bottom up expansion, growing from $\chi_{_{4N:4N}}$ (a chain with a single site at position $1$) to $\chi_{_{1:4N}}$. 
Here, unless stated otherwise, we follow the top down expansion i), through which
\eqref{eq:chissh4general} is expanded from the bottom corner to read as
\begin{equation}
\chi_{_{1:4N}}(\lambda)=\lambda\chi_{_{1:4N-1}}(\lambda)-t_1^2\chi_{_{1:4N-2}}(\lambda).
\label{eq:chissh4}
\end{equation}
However, different relations hold for $\chi_{_{1:4N-1}}$,$\chi_{_{1:4N-2}}$ and $\chi_{_{1:4N-3}}$, as the hopping parameter at the last term of (\ref{eq:chissh4}) is changed to $t_2$, $t_3$ and $t_4$, respectively, before returning to $t_1$ again. 
Therefore we write (\ref{eq:chissh4}) as a system of coupled recurrence relations. Defining $\chi_{_{1:4n+1-i}}(\lambda)=\chi_n^i(\lambda)$, with $i=1,2,3,4$ and $n=0,1,2,\dots,N$ (where $\chi_0^i(\lambda)$ will be determined by the boundary conditions), we get
\begin{eqnarray}
\chi_n^1(\lambda)&=&\lambda\chi_n^2(\lambda)-t_1^2\chi_n^3(\lambda),
\label{eq:chissh41}
\\
\chi_n^2(\lambda)&=&\lambda\chi_n^3(\lambda)-t_2^2\chi_n^4(\lambda),
\label{eq:chissh42}
\\
\chi_n^3(\lambda)&=&\lambda\chi_n^4(\lambda)-t_3^2\chi_{n-1}^1(\lambda),
\label{eq:chissh43}
\\
\chi_n^4(\lambda)&=&\lambda\chi_{n-1}^1(\lambda)-t_4^2\chi_{n-1}^2(\lambda).
\label{eq:chissh44}
\end{eqnarray}
Using (\ref{eq:chissh42}-\ref{eq:chissh44}) to develop (\ref{eq:chissh41}) we arrive, after some algebra, at
\begin{eqnarray}
\chi_n^1(\lambda)&=&\big[\lambda^4-(t_1^2+t_2^2+t_3^2+t_4^2)\lambda^2+ t_1^2t_3^2+t_2^2t_4^2\big]\chi_{n-1}^1(\lambda) \nonumber
\\
&-&(t_1t_2t_3t_4)^2\chi_{n-2}^1(\lambda).
\label{eq:chi1ssh4a}
\end{eqnarray}

Now, our strategy will be to identify the $\lambda$ parameter with one of the energy bands in $k$-space of the SSH$_4$ model for an infinite chain, which can be straightforwardly found to be given by
\begin{widetext}
\begin{equation}
\lambda_{\pm\pm}(\cos k)=\pm\frac{1}{\sqrt{2}}\sqrt{t_1^2+t_2^2+t_3^2+t_4^2\pm\sqrt{(t_1^2+t_2^2+t_3^2+t_4^2)^2-4(t_1^2t_3^2+t_2^2t_4^2-2t_1t_2t_3t_4\cos k)}},
\label{eq:ebandsssh4}
\end{equation}
\end{widetext}
where $k\in[-\pi,\pi[$ and the lattice spacing was set to $a=1$.
In other words, we are searching for the eigenenergies of the open SSH$_4$ model within the energy range of each band of the spectrum. Possible edge states, such as topological edge states which appear in some energy gap, fall outside the parametrization ranges and have to be dealt separately, as we will show later on.
The following relation holds for all bands in (\ref{eq:ebandsssh4}),
\begin{equation}
\lambda^4-(t_1^2+t_2^2+t_3^2+t_4^2)\lambda^2+ t_1^2t_3^2+t_2^2t_4^2=2t_1t_2t_3t_4\cos k,
\label{eq:coskrelationssh4}
\end{equation}
which in turn simplifies (\ref{eq:chi1ssh4a}) to
\begin{equation}
\chi_n^1(\lambda)=2t_1t_2t_3t_4\cos k\chi_{n-1}^1 - (t_1t_2t_3t_4)^2\chi_{n-2}^1(\lambda).
\label{eq:recurrencechissh4}
\end{equation}
All $\chi_{n\geq 2}^1(\lambda)$ can then be obtained from the boundary conditions $\chi_0^1(\lambda)$ and $\chi_1^1(\lambda)$.
We set $\chi_0^1(\lambda)=1$ and determine $\chi_1^1(\lambda)$ from (\ref{eq:chissh4general}),
\begin{eqnarray}
\chi_1^1(\lambda)&=&\chi_{_{1:4}}(\lambda)=
\begin{vmatrix}
\lambda&t_3&0&0
\\
t_3&\lambda&t_2&0
\\
0&t_2&\lambda&t_1
\\
0&0&t_1&\lambda
\end{vmatrix} \nonumber
\\
&=&2t_1t_2t_3t_4\cos k+t_4^2(\lambda^2-t_2^2),
\label{eq:chi11ssh4a}
\end{eqnarray}
where (\ref{eq:coskrelationssh4}) was used again in the last step.
By defining $\chi_{1K}^1(\lambda)$ as the determinant of the kernel of $\chi_1^1(\lambda)$, that is, $\chi_{1K}^1(\lambda)$ is constructed by taking the first and last rows and columns of $\chi_1^1(\lambda)$,
\begin{equation}
\chi_{1K}^1(\lambda)=\begin{vmatrix}
\lambda&t_2\\
t_2&\lambda
\end{vmatrix}
=\lambda^2-t_2^2,
\end{equation}
then (\ref{eq:chi11ssh4a}) can be simplified to
\begin{equation}
\chi_1^1(\lambda)=2t_1t_2t_3t_4\cos k+t_4^2\chi_{1K}^1(\lambda).
\label{eq:chi11ssh4b}
\end{equation}

The recurrence relation in (\ref{eq:recurrencechissh4}) can be further simplified by defining
\begin{equation}
W_n^1(\lambda,\cos k):=(t_1t_2t_3t_4)^{N-n}\chi_n^1(\lambda),
\label{eq:wchirelationssh4}
\end{equation}
where in turn $\lambda=\lambda(\cos k)$, becoming then
\begin{equation}
W_n^1=2\cos kW_{n-1}^1-W_{n-2}^1,
\end{equation}
where the dependence on $\lambda$ and $\cos k$ was left implied for convenience.
These $W_n^1$ follow the same recurrence relation as the Chebyshev polynomials of the second-kind $U_n=U_n(\cos k)$, but with modified boundary conditions, in relation to $U_{-1}=0$, $U_0=1$ and $U_1=2\cos k$.
With $\chi_0^1(\lambda)=1$ and $\chi_1^1(\lambda)$ given in (\ref{eq:chi11ssh4b}), we can use (\ref{eq:wchirelationssh4}) to determine the boundary conditions $W_0^1$ and $W_1^1$,
\begin{eqnarray}
W_0^1&=&(t_1t_2t_3t_4)^{N}(1+0) \nonumber
\\
&=&(t_1t_2t_3t_4)^{N}(U_0+U_{-1}),
\label{eq:w0ssh4}
\\
W_1^1&=&(t_1t_2t_3t_4)^{N}\big(2\cos k + \alpha_1(\lambda)\big) \nonumber
\\
&=&(t_1t_2t_3t_4)^{N}\big(U_1 + \alpha_1(\lambda)U_0\big),
\label{eq:w1ssh4}
\end{eqnarray}
with $\alpha_1(\lambda)=\frac{t_4}{t_1t_2t_3}\chi_{1K}^1(\lambda)$.
Comparing (\ref{eq:w0ssh4}) and (\ref{eq:w1ssh4}), the general relation for $W_n^1$, with $n\geq 2$, can be readily found to yield
\begin{equation}
W_n^1(\lambda,\cos k)=(t_1t_2t_3t_4)^{N}\big[U_n(\cos k)+\alpha_1(\lambda)U_{n-1}(\cos k)\big].
\label{eq:modchebyshevssh4}
\end{equation}

\begin{figure}[h]
	\begin{center}
		\includegraphics[width=0.49 \textwidth]{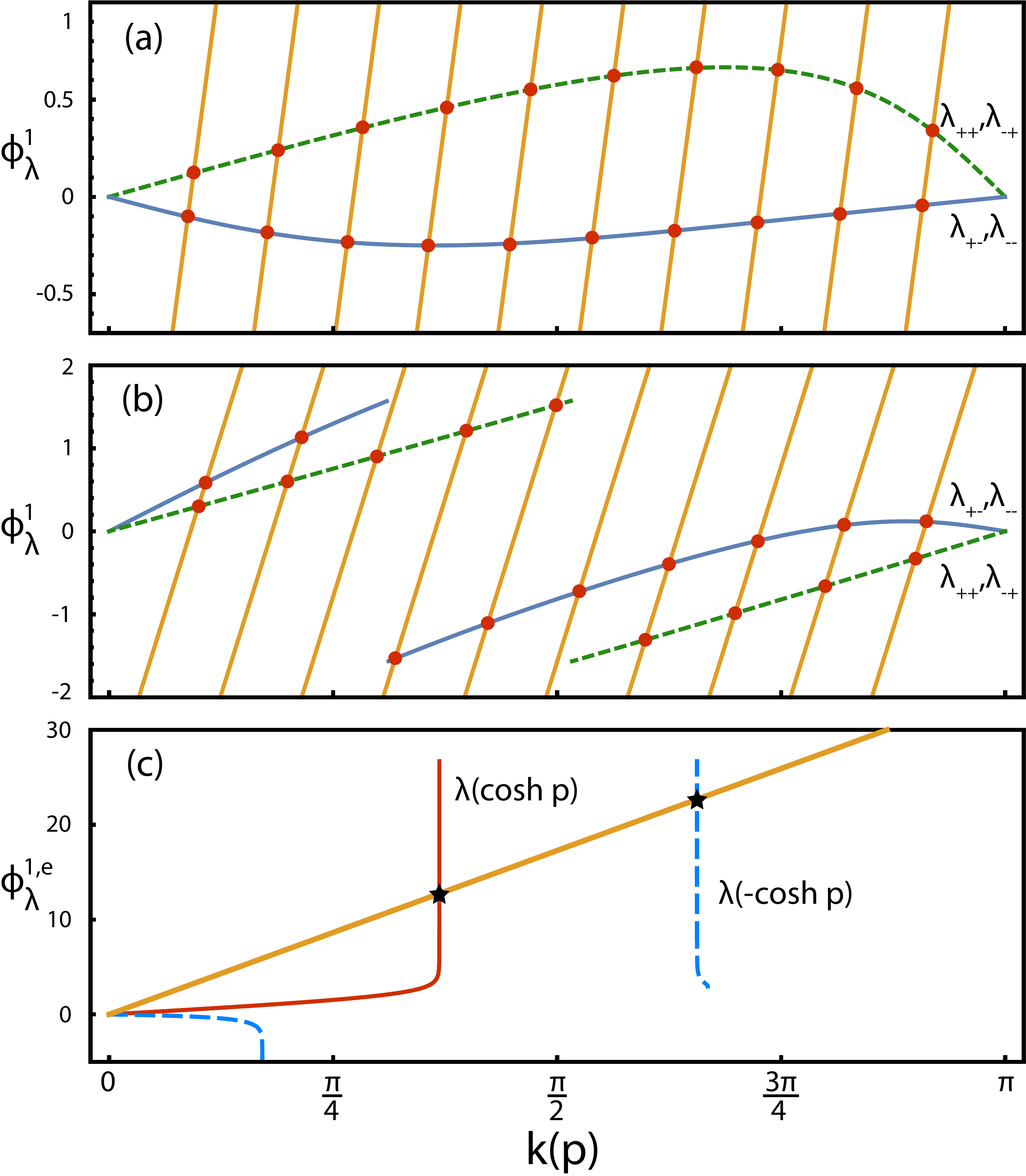}
	\end{center}
	\caption{(a) Momentum shift $\phi_\lambda^1$ as a function of $k$ for the SSH$_4$ model with $\delta=\frac{\pi}{2}$ [see (\ref{eq:deltapiovertwo})]. Equally spaced orange lines represent the successive  $f_n(k)=(N+1)k-n\pi$, with $n=1,2,\dots,N$ and $N=10$ the number of unit cells. The $k_n$ values of the red dots at the intersections are the momentum values of the eigenstates.
	(b) Same as in (a) but with inverted hopping parameters $\delta=0$ [see (\ref{eq:deltazero})].
Note that each band has one less $k$-solution than in (a).
(c) Imaginary momentum shift $\phi_\lambda^{1,e}$ as a function of $p$ for the SSH$_4$ model with the parameters of (b). 
Only regions where $\phi_\lambda^{1,e}(p)$ is real are depicted.
Orange line represents $f^{e}(p)=(N+1)p$, where $N=10$ is the number of unit cells.
Only the degenerate edge phases of the chiral pair of bands $\lambda_{+-}$ and $\lambda_{--}$ are represented here, since only they have non-trivial $p$ solutions, given by the $p$ values of the black stars at the intersections.}
	\label{fig:phasessh4}
\end{figure}
For each band $\lambda$, the corresponding $k$ solutions are found by solving the characteristic equation $\chi_N^1(\lambda)=0$.
Noting that $\chi_N^1(\lambda)=W_N^1(\lambda,\cos k)$, through (\ref{eq:wchirelationssh4}), and using the well known result
\begin{equation}
U_n(\cos k)=\frac{\sin\big[(n+1)k\big]}{\sin k},
\label{eq:chebyshevcosk}
\end{equation}
the characteristic equation $W_N^1(\lambda,\cos k)=0$ can be manipulated to read, using standard trigonometric identities, as
\begin{equation}
\cot\big[(N+1)k\big]=\frac{1}{\alpha_1(\lambda)\sin k} + \cot k,
\label{eq:cotssh4}
\end{equation}
from where one finally arrives at
\begin{eqnarray}
\phi_\lambda^1(k)&=&(N+1)k-n\pi,\ \ \ n=1,2,\dots,N,
\label{eq:ksolutionsssh4}
\\
\phi_\lambda^1(k)&=&\cot^{-1}\Big[\frac{1}{\alpha_1(\lambda)\sin k}+\cot k\Big], 
\label{eq:phissh4}
\end{eqnarray} 
where the phase, defined in the interval $\phi_\lambda^1(k)\in ]-\frac{\pi}{2},\frac{\pi}{2}]$, represents the momentum shift in relation to the usual $\phi_\lambda^1(k)=0$ case, for which one recovers $k_n=\frac{n\pi}{N+1}$.
For every band $\lambda$ we solve (\ref{eq:ksolutionsssh4}) for each $n$ to find the set of allowed $k_n$ values within the Reduced Brillouin Zone (RBZ), $k_n\in[0,\pi[$.
An example of the geometrical determination of the $k$ states, for a system with $N=10$ unit cells and $\delta=\frac{\pi}{2}$, is shown in Fig.~\ref{fig:phasessh4}(a).
The energy of these $k$-states of the open chain is given by the corresponding value of $\lambda(k)$.
Each of the two distinct $\phi_\lambda^1(k)$ is twice degenerate, since the SSH$_4$ model is bipartite and, therefore, has chiral symmetry defined as $CH(k)C^{-1}=-H(k)$, so that the $\lambda$ bands come in chiral pairs sharing the same $\phi_\lambda^1(k)$ and the same set of $\{k_n\}$.
In general, the SSH$_m$ model has $m$ distinct $\phi_\lambda^1(k)$ for $m$ odd, and $m/2$ distinct $\phi_\lambda^1(k)$ for $m$ even.
These results lead to two important remarks: i) defined in the RBZ, the absolute momentum $k$ is still a good quantum-number, and ii) contrary to periodic models, the set of allowed $k$-values can, in principle, be different for every band. 

\begin{figure}[h]
	\begin{center}
		\includegraphics[width=0.49 \textwidth]{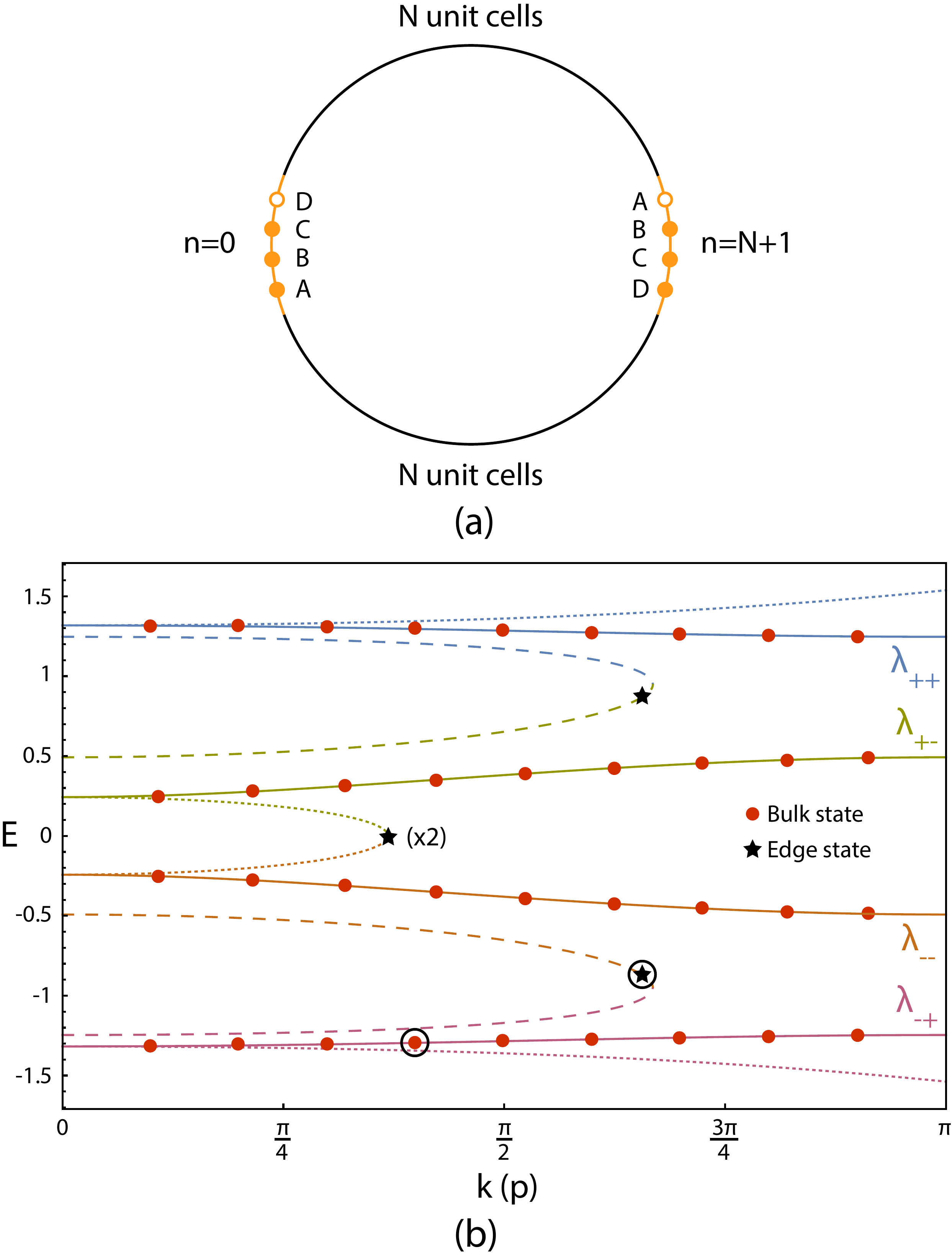}
	\end{center}
	\caption{(a) Periodic SSH$_4$ model with $2N+2$ unit cells. Sites of unit cells $n=0,N+1$ are highlighted. By combining symmetric $k$-states and imposing nodes at $\ket{0,D}$ and $\ket{N+1,A}$, an open SSH$_4$ chain is created at the upper $N$ unit cells.
		(b) Energy spectrum of the SSH$_4$ model with $\delta=0$ [see (\ref{eq:deltazero})] in the RBZ as a function of $k$ for the $\lambda(\cos k)$ bands (solid curves) and as a function of $p$ for the $\lambda(\cosh p)$ bands (dotted curves) and $\lambda(-\cosh p)$ bands (dashed curves).
		Identically colored bands give the $\cos k$ and $\pm\cosh p$ parametrizations of the same $\lambda$ band.
		The $k$ ($p$) solutions for an open chain with $N=10$ unit cells are indicated by the red dots (black stars).
		Highlighted encircled states correspond to those of Fig.~\ref{fig:eigvecsssh4}.}
	\label{fig:espectrumssh4}
\end{figure}

Having determined the absolute momentum $k_n$ and respective energy of all eigenstates, we want to find now the spatial profile of these states along the open chain with $N$ unit cells.
The treatment followed here consists of assuming a larger periodic system (we consider $2N+2$ unit cells to simplify, but the same procedure holds for periodic chains with $n\geq N+2$ unit cells) and then combine degenerate $k$-states in order to impose nodes at specific positions, such that an open chain of $N$ unit cells, with eigenstates satisfying the OBC, can be extracted from the full periodic chain.
A general $k$-state of the periodic SSH$_4$ model with $2N+2$ unit cells of Fig.~\ref{fig:espectrumssh4}(a) can be written as
\begin{eqnarray}
\ket{\varphi_\lambda(k)}=\frac{1}{\sqrt{2N+2}}&\sum\limits_{j=0}^{2N+1}&e^{ikj}
\begin{bmatrix}
a_\lambda(k)e^{-i\theta_\lambda^a(k)}
\\
b_\lambda(k)e^{-i\theta_\lambda^b(k)}
\\
c_\lambda(k)e^{-i\theta_\lambda^c(k)}
\\
d_\lambda(k)
\end{bmatrix},
\label{eq:eigstatessh4pbc}
\\
&\sum\limits_{\eta=a,b,c,d}&\vert\eta_\lambda(k)\vert^2=1
\end{eqnarray}
where $k\in[-\pi,\pi[$, the phase of the $D$-component was set to zero for convenience and $\eta_\lambda(k)\in\mathbb{R}^+_0$, with $\eta=A,B,C,D$.
From the presence of time-reversal symmetry it follows that $\theta_\lambda^\eta(k)=-\theta_\lambda^\eta(-k)$ and $\eta(k)=\eta(-k)$.
The eigenfunctions of the open chain can be found through the standard combination of degenerate symmetric $k$-states of the periodic chain,
\begin{eqnarray}
\ket{\psi_\lambda(k)}&=&\frac{1}{\sqrt{2}}\big(\ket{\varphi_\lambda(k)}-\ket{\varphi_\lambda(-k)}\big) \nonumber
\\
&=&\frac{1}{\sqrt{N+1}}\sum\limits_{j=0}^{2N+1}\ket{u_{\lambda,j}(k)},
\label{eq:eigstatessh4antisym}
\\
\ket{u_{\lambda,j}(k)}&=&
\begin{bmatrix}
a_\lambda(k)\sin[kj-\theta_\lambda^a(k)]
\\
b_\lambda(k)\sin[kj-\theta_\lambda^b(k)]
\\
c_\lambda(k)\sin[kj-\theta_\lambda^c(k)]
\\
d_\lambda(k)\sin[kj]
\end{bmatrix},
\label{eq:blochstatessh4}
\end{eqnarray}
where $k$ is now within the RBZ.
By identifying $\ket{j,\eta}$ as the $\eta=A,B,C,D$ component of $\ket{u_{\lambda,j}(k)}$, the boundary conditions are defined as $\braket{0,D}{\psi_\lambda(k)}=0$, which is automatically satisfied, and $\braket{N+1,A}{\psi_\lambda(k)}=0$, which yields (\ref{eq:ksolutionsssh4}) since we can directly identify $\theta_\lambda^a(k)\equiv\phi_\lambda^1(k)$.
We decompose $\ket{\psi_\lambda(k)}$ in two terms,
\begin{equation}
\ket{\psi_\lambda(k)}=\frac{1}{\sqrt{N+1}}\Big[\sum\limits_{j=1}^N\ket{u_{\lambda,j}(k)}+\sum\limits_{j=N+1}^{2N+2}\ket{u_{\lambda,j}(k)}\Big],
\label{eq:eigstatetotalchainssh4}
\end{equation}
where $\ket{2N+2,\eta}\equiv\ket{0,\eta}$.
These two terms are isolated from one another due to the nodes at $\ket{0,D}$ and $\ket{N+1,A}$.
Finally, to get the form of an eigenstate of our open SSH$_4$ chain with $N$ unit cells we drop the second term at the right-and side of (\ref{eq:eigstatetotalchainssh4}) and re-normalize our state,
\begin{eqnarray}
\ket{\psi_\lambda(k)}&=&\zeta_\lambda(k)\sqrt{\frac{2}{N+1}}\sum\limits_{j=1}^N\ket{u_{\lambda,j}(k)},
\label{eq:eigstatessh4}
\\
\vert \zeta_\lambda(k)\vert^2&=&\frac{N+1}{2}\Big[\sum\limits_{j=1}^N\vert u_{\lambda,j}(k)\vert^2\Big]^{-1},
\label{eq:normfactorssh4}
\end{eqnarray}
where, in general, $\vert \zeta_\lambda(k)\vert\neq 1$, as can be expected from the different sizes of the upper and lower chains separated by the nodes at $\ket{0,D}$ and $\ket{N+1,A}$ in Fig.~\ref{fig:espectrumssh4}(a). 
We presented this detailed derivation of the eigenstates under OBC in order to show that, contrary to what is sometimes assumed \cite{Delplace2011}, one cannot directly extrapolate from the well known results for the chain with a single hopping parameter and set $\zeta_\lambda(k)=1$.
Also, it is important to notice that since, as mentioned above, under OBC the set of allowed $k$-values can be different for every band, it is clear that they do not in general coincide with the allowed $k$-values under PBC. 
In other words, the degenerate symmetric $k$-states of the form of \eqref{eq:eigstatessh4pbc} that are being combined to produce nodes at specific positions \textit{are not} eigenstates of the larger chain with PBC. 

\subsection{Edge states}
\label{subsec:outofbandssh4}
For the hopping parameters considered in Fig.~\ref{fig:phasessh4}(b), the parametrization of $\phi_\lambda^1(k)$ using each of the $\lambda$ bands finds one less $k_n$ than for the set of parameters used in Fig.~\ref{fig:phasessh4}(a).
Since the total number of states is fixed to $4N$, this implies the existence of four edge states that drifted away from the bulk bands and into the energy gaps as the $t_i$'s are varied adiabatically from those of Fig.~\ref{fig:phasessh4}(a) to those of Fig.~\ref{fig:phasessh4}(b), which, in turn, implies that a transition takes place between these two cases (with and without edge states).
In order to find the edge states we therefore have to extend the energy range of the parametrization, which can be achieved by considering a complex $k=q+ip$ \cite{Delplace2011,Banchi2013,Hugel2014,Marques2017,Duncan2018}.
The imaginary part $p$ of the complex momentum is the inverse localization length of the edge state.
The condition of keeping all $\lambda(k)$ real imposes that $q=0\vee\pi$ (the ``+" and ``-" solutions, respectively), such that now we have $\lambda(\cos k)\to\lambda(\pm\cosh p)$. 
As can be seen in Fig.~\ref{fig:espectrumssh4}(b), the $\lambda(\pm\cosh p)$ bands fill all the energy gaps between the $\lambda(\cos k)$ bands, so that each in-gap edge state falls into the energy range of its corresponding $\lambda(\pm\cosh p)$ band.
The relation in (\ref{eq:chebyshevcosk}) now becomes
\begin{equation}
U_n(\pm\cosh p)=(\pm)^n\frac{\sinh [(n+1)p]}{\sinh p}.
\label{eq:chebyshevcoshk}
\end{equation}
With the substitution $\cos k\to\pm\cosh p$ in (\ref{eq:modchebyshevssh4}), the characteristic equation $W_N^1(\lambda,\pm\cosh p)=0$ yields
\begin{eqnarray}
\phi_{\lambda,\pm}^{1,e}(p)&=&(N+1)p,
\label{eq:psolutionsssh4}
\\
\phi_{\lambda,\pm}^{1,e}(p)&=&\coth^{-1}\Big[\frac{1}{\pm\alpha^{e}_{1,\pm}(\lambda)\sinh p}+\coth p\Big],
\label{eq:phiobssh4}
\end{eqnarray} 
where the \textbf{e}dge $\alpha^{1,e}_{1,\pm}(\lambda)$ is given by applying $\cos k\to\pm\cosh p$ to the $\alpha_1(\lambda)$ defined in (\ref{eq:w1ssh4}) \footnote{It can be easily shown that equations (5) and (10) in Ref.~[\onlinecite{Duncan2018}] can be reduced to our equations (\ref{eq:ksolutionsssh4}) and (\ref{eq:psolutionsssh4}), respectively. Here, we further determine the general expression for $\phi_\lambda^1(k)$ and $\phi_{\lambda,\pm}^{1,e}(p)$ [see (\ref{eq:phissh4}) and (\ref{eq:phiobssh4})].} and $\phi_{\lambda,\pm}^{1,e}(p)$ represents the imaginary momentum shift from the $k=0,\pi$ states.
Note that, contrary to the $N$ bulk equations that have to be solved for each $\lambda$ band [see \eqref{eq:ksolutionsssh4}], there is only one edge equation for each band, even though it can in general have multiple solutions, that is, multiple edge states belonging to the same edge band.
For the hopping parameters of Fig.~\ref{fig:phasessh4}(b) the only bands with non-trivial $p$-solutions are $\lambda_{--}(\pm\cosh p)$ and $\lambda_{+-}(\pm\cosh p)$.
Since these middle bands form a chiral pair they share the same $\phi_{\lambda,\pm}^{1,e}(p)$ and, therefore, the same $p$-solutions, as depicted in Fig.~\ref{fig:phasessh4}(c).
By substituting the $k$-solutions of Fig.~\ref{fig:phasessh4}(b) and the $p$-solutions of Fig.~\ref{fig:phasessh4}(c) in their respective energy bands, the full energy spectrum can be found, as shown in Fig.~\ref{fig:espectrumssh4}(b).
There is a duality between the set of $k$-bands and the set of $p$-bands, in the sense that each of them exactly fills the energy gaps of the other.

\begin{figure}[h]
	\begin{center}
		\includegraphics[width=0.47 \textwidth]{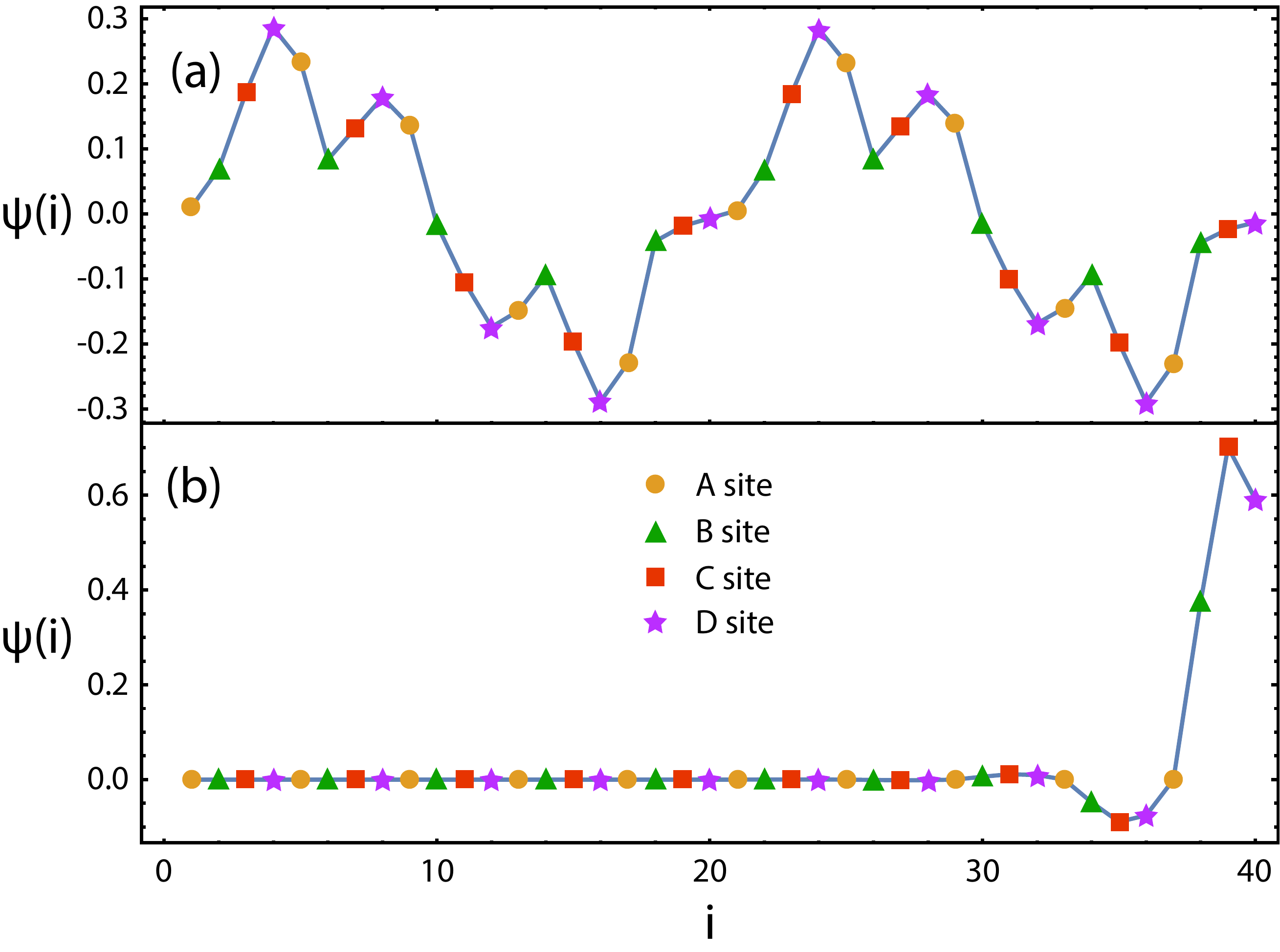}
	\end{center}
	\caption{Spatial profile along the $i$ sites of an open SSH$_4$ chain with $N=10$ unit cells of the wavefunction of the (a) bulk and (b) edge states highlighted in Fig.~\ref{fig:espectrumssh4}(b) computed through (\ref{eq:eigstatessh4}) and (\ref{eq:eigstateobssh4}), respectively. 
	Agreement with exact diagonalization results was verified.} 
	\label{fig:eigvecsssh4}
\end{figure}

Since $\phi_{\lambda,\pm}^{1,e}(p)$ in (\ref{eq:phiobssh4}) has to follow directly from $\phi_\lambda^1(k)$ in (\ref{eq:phissh4}) after substituting $k\to q+ip$, with $q=0\vee\pi$, we find that $\theta_\lambda^a(k)=\phi_\lambda^1(k)\to i\phi_{\lambda,\pm}^{1,e}(p)$, with equivalent relations holding for $\theta_\lambda^b(k)\to i\theta_{\lambda,\pm}^{b,e}(p)$ and $\theta_\lambda^c(k)\to i\theta_{\lambda,\pm}^{c,e}(p)$ in (\ref{eq:blochstatessh4}), as will be shown in Section~\ref{sec:isshml1}.
By further applying $\eta_\lambda(k)\to \eta_{\lambda,\pm}^{e}(p)$ to (\ref{eq:blochstatessh4}), with $\eta=a,b,c,d$, the eigenstates of the edge states are written as, apart from a global phase factor,
\begin{eqnarray}
\ket{\psi^{e}_ {\lambda,\pm}(p)}&=&\zeta^e_{\lambda,\pm}(p)\sum\limits_{j=1}^N(\pm)^j\ket{u^{e}_ {\lambda,j,\pm}(p)},
\label{eq:eigstateobssh4}
\\
\ket{u^{e}_ {\lambda,j,\pm}(p)}&=&
\begin{bmatrix}
a^{e}_{\lambda,\pm}(p)\sinh[pj-\phi_{\lambda,\pm}^{1,e}(p)]
\\
b^{e}_{\lambda,\pm}(p)\sinh[pj-\theta_{\lambda,\pm}^{b,e}(p)]
\\
c^{e}_{\lambda,\pm}(p)\sinh[pj-\theta_{\lambda,\pm}^{c,e}(p)]
\\
d^{e}_{\lambda,\pm}(p)\sinh[pj]
\end{bmatrix},
\label{eq:blochstateobssh4}
\\
\vert \zeta^{e}_{\lambda,\pm}(p)\vert^2&=&\Big[\sum\limits_{j=1}^N\vert u^{e}_ {\lambda,j,\pm}\vert^2\Big]^{-1},
\label{eq:normobssh4}
\end{eqnarray}
where we highlighted that $\theta_{\lambda,\pm}^{a,e}(p)\equiv\phi_{\lambda,\pm}^{1,e}(p)$.
These eigenstates are edge localized.
Examples of a bulk state and a right-edge localized state computed using \eqref{eq:eigstatessh4} and \eqref{eq:eigstateobssh4}, respectively, are shown in Fig.~\ref{fig:eigvecsssh4}.

\section{ISSH model}
\label{sec:issh}

In order to see the effect of introducing arbitrary on-site potentials within the unit cell let us study the ISSH model under OBC.
Its Hamiltonian can be written as
\setcounter{MaxMatrixCols}{20}
\begin{eqnarray}
H&=&-\begin{pmatrix}
-v_2&t_1&&&
\\
t_1&-v_1&t_2&
\\
&t_2&-v_2&&
\\
&&\ddots&\ddots&\ddots
\\
&&&&-v_1&t_2
\\
&&&&t_2&-v_2&t_1
\\
&&&&&t_1&-v_1
\end{pmatrix},
\\
H'&=&\frac{H-v_1I}{t_1}=-
\begin{pmatrix}
-v&1&&&
\\
1&0&t&
\\
&t&-v&&
\\
&&\ddots&\ddots&\ddots
\\
&&&&0&t
\\
&&&&t&-v&1
\\
&&&&&1&0
\end{pmatrix},
\end{eqnarray}
where $I$ is the identity matrix, $t=t_2/t_1$ and $v=(v_2-v_1)/t_1$. 
Note that we always have the freedom of setting one hopping parameter to one (the energy unit) and one on-site potential as the zero potential energy level.
Since we have two sites per unit cell and $N$ unit cells, we get a system of two coupled recurrence relations for the characteristic polynomials,
\begin{eqnarray}
\chi_n^1(\lambda)&=&\lambda\chi_n^2(\lambda)-\chi_{n-1}^1(\lambda),
\label{eq:chissh1}
\\
\chi_n^2(\lambda)&=&(\lambda-v)\chi_{n-1}^1(\lambda)-t^2\chi_{n-1}^2(\lambda),
\label{eq:chissh2}
\end{eqnarray}
for $n=0,1,\dots,N$.
Using (\ref{eq:chissh2}) to develop (\ref{eq:chissh1}) we arrive at
\begin{equation}
\chi_n^1(\lambda)=\big[\lambda^2-v\lambda-1-t^2\big]\chi_{n-1}^1(\lambda)-t^2\chi_{n-2}^1(\lambda).
\label{eq:chi1ssha}
\end{equation}
\begin{figure}[h]
	\begin{center}
		\includegraphics[width=0.49 \textwidth]{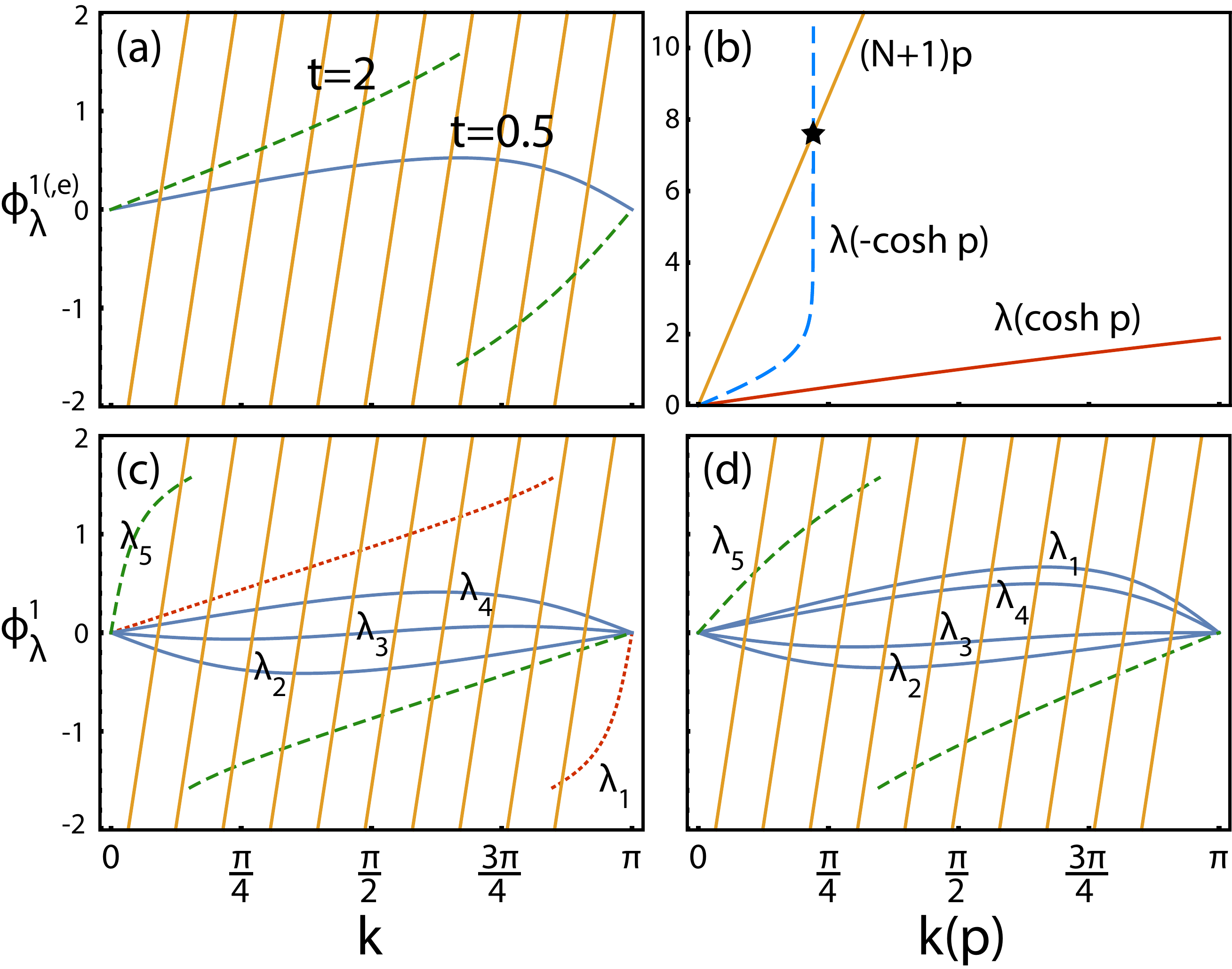}
	\end{center}
	\caption{(a) Degenerate $\phi_\lambda^1$ phases of $\lambda_{+}$ and $\lambda_{-}$ bands as a function of $k$ for the ISSH model with arbitrary $v$ and two different $t$ values. Equally spaced orange lines represent the successive  $f_n(k)=(N+1)k-n\pi$, with $n=1,2,\dots,N$ and $N=10$ the number of unit cells. The $k_n$ values at the intersections are the eigenstates.
		(b) Degenerate $\phi_\lambda^{1,e}$ phases of $\lambda_{+}$ and $\lambda_{-}$ bands as a function of $p$ for the ISSH with arbitrary $v$ and $t=2$. 
		Orange line represents $f^{e}(p)=(N+1)p$, where $N=10$ is the number of unit cells.
		The non-trivial $p$-solutions come from the intersections of the $\lambda(-\cosh p)$ bands.
		(c) and (d) Same as in (a) but for the SSH$_5$ model  with $(t_1,t_2,t_3,t_4,t_5)=(1,0.8,0.6,0.4,0.2)$ (that is, $\delta=\frac{\pi}{2}$) and for the ISSH$_5$ with the same hopping parameters and $v_i=t_i$, respectively.
		The labeling of the $\lambda_i$ bands follows an increasing energy order.
		Non-solid curves have one less solution than the solid curves.}
	\label{fig:phasessh5}
\end{figure}
The energy bands of the periodic model are given by
\begin{equation}
\lambda_\pm(k)=\frac{1}{2}\Big(v\pm\sqrt{v^2+4(1+t^2+2t\cos k)}\Big),
\end{equation}
with lattice spacing $a=1$, from where both bands can be found to obey the following relation,
\begin{equation}
\lambda^2-v\lambda-1-t^2=2t\cos k,
\end{equation}
which, when inserted back in (\ref{eq:chi1ssha}), yields a relation equivalent to that of (\ref{eq:recurrencechissh4}), 
\begin{equation}
\chi_n^1(\lambda)=2t\cos k\chi_{n-1}^1 - t^2\chi_{n-2}^1(\lambda),
\label{eq:recurrencechissh}
\end{equation}
from where one can follow the same procedure as for the SSH$_4$ model to arrive at the same expressions for $\phi_\lambda^1(k)$ and $\phi_{\lambda,\pm}^{1,e}(p)$, with $\alpha_1(\lambda)=t\chi_{1K}^1(\lambda)$, showing them to be insensitive to the introduction of the on-site potential $v$.
From
\begin{equation}
\chi_1^1(\lambda)=\chi_{_{1:2}}(\lambda)=
\begin{vmatrix}
\lambda-v&1
\\
1&\lambda
\end{vmatrix},
\end{equation}
we find $\chi_{1K}^1(\lambda)=1$ and $\alpha_1(\lambda)=t$, in accordance with [\onlinecite{Delplace2011}].
Only for the ISSH model, the simplest of the ISSH$_m$ models, is $\chi_{1K}^1(\lambda)$ [and therefore $\alpha_1(\lambda)$] also independent of any on-site potentials. 
As such, even though a finite $v$ breaks chiral symmetry in the ISSH model, the $\phi_\lambda^1(k)$ and $\phi_{\lambda,\pm}^{1,e}(p)$ phases of both bands remain the same for all $v$, as can be seen in Fig.~\ref{fig:phasessh5}(a) and Fig.~\ref{fig:phasessh5}(b), respectively.
The edge states are shown in Fig.~\ref{fig:phasessh5}(b) to be in the $\lambda(-\cosh p)$ bands, \textit{i.e.}, the real part of their momentum is $q=\pi$, which is the gap closing point at $t=1$ in the thermodynamic limit.

In the case of the ISSH$_5$ model, for instance, we have a $v_i$ sensitive $\alpha_1(\lambda)$, given in this case by
\begin{equation}
\alpha_1(\lambda)=-\frac{t_5}{t_1t_2t_3t_4}\chi_{1K}^1,
\label{eq:alphaissh5}
\end{equation}
with
\begin{equation}
\chi_{1K}^1(\lambda)=
\begin{vmatrix}
\lambda-v_4&t_3&0
\\
t_3&\lambda-v_3&t_2
\\
0&t_2&\lambda-v_2
\end{vmatrix}.
\label{eq:chi1k1issh5}
\end{equation}
where the $\lambda$ bands depend on all $t_i$ and $v_i$.
Given that the $\phi_\lambda^1(k)$ in (\ref{eq:phissh4}) depend on the set of on-site potentials $\{v_i\}$, the corresponding set of $k$-solutions will also change with $\{v_i\}$, as can be seen by comparing the solutions of the SSH$_5$ model in Fig.~\ref{fig:phasessh5}(c) with those of the ISSH$_5$ model in Fig.~\ref{fig:phasessh5}(d).
In particular, qualitatively different behavior between these two cases is found for $\lambda_1$, having one less $k$-solution in the SSH$_5$ model than for the ISSH$_5$ model, that is, one of the edge states of SSH$_5$ model becomes a bulk state in the ISSH$_5$ model.

\section{General method}
\label{sec:genmet}
In this section we outline a summarized and operative version of the method for finding the eigenstates of a general ISSH$_m$ model under OBC, with the unit cell of Fig.~\ref{fig:ionicsshm}, omitting some intermediate steps explicitly shown in the previous sections.
\begin{enumerate}[leftmargin=*]
	\item First one starts by computing the $\lambda(\cos k)$ energy bands  under PBC. 
	
	\item The system of coupled recurrence relations for the characteristic polynomials $\chi_n^j(\lambda)$, with $j=1,2,\dots,m$ and $n=1,2,\dots,N$, where $N$ is the number of unit cells under OBC, can be written as
	\begin{eqnarray}
	\chi_n^i(\lambda)&=&(\lambda-v_i)\chi_n^{i+1}(\lambda)-t_i^2\chi_n^{i+2}(\lambda),
	\label{eq:chiissh1}
	\\
	\vdots \nonumber
	\\
	\chi_n^{m-1}(\lambda)&=&(\lambda-v_{m-1})\chi_n^{m}(\lambda)-t_{m-1}^2\chi_{n-1}^1(\lambda),
	\label{eq:chiisshmminus1}
	\\
	\chi_n^{m}(\lambda)&=&(\lambda-v_{m})\chi_{n-1}^1(\lambda)-t_{m}^2\chi_{n-1}^2(\lambda),
	\label{eq:chiisshm}
	\end{eqnarray}
	where $i=1,2,\dots,m-2$. 
	Using these equations and the expressions for the $\lambda$ bands to develop $\chi_n^1$ one arrives at
	\begin{eqnarray}
	\chi_n^1(\lambda)&=&2T\cos k\chi_{n-1}^1(\lambda)-T^2\chi_{n-2}^1(\lambda),
	\label{eq:chi1isshm}
	\\
	T&=&(-1)^m\prod\limits_{j=1}^mt_j,
	\label{eq:tprod}
	\end{eqnarray}
	where the pre-factor to the product operator comes from the ``-" sign of the convention we adopted in the definition of the hopping parameters at the Hamiltonian [see (\ref{eq:hamiltssh4})].
	The boundary conditions to (\ref{eq:chi1isshm}) are given by
	\begin{widetext}
	\begin{eqnarray}
	\chi_0^1(\lambda)&=&1,
	\\
	\chi_1^1(\lambda)&=&
	\begin{vmatrix}
	(\lambda-v_m)&t_{m-1}&&&
	\\
	t_{m-1}&(\lambda-v_{m-1})&t_{m-2}&
	\\
	&t_{m-2}&(\lambda-v_{m-2})&&
	\\
	&&\ddots&\ddots&\ddots
	\\
	&&&&(\lambda-v_3)&t_2
	\\
	&&&&t_2&(\lambda-v_2)&t_1
	\\
	&&&&&t_1&(\lambda-v_1)
	\end{vmatrix}.
	\label{eq:chi11isshm}
	\end{eqnarray}
    \end{widetext}

    \item
    The characteristic polynomial $\chi_n^1(\lambda)$ can be recast as
    \begin{eqnarray}
    W_n^1(\lambda,\cos k)&:=&T^{N-n}\chi_n^1(\lambda),
    \label{eq:chi1nisshm}
    \\
    W_n^1(\lambda,\cos k)&=&T^{N}\big[U_n(\cos k)+\alpha_1(\lambda)U_{n-1}(\cos k)\big],
    \label{eq:modchebyshevisshm}   
    \end{eqnarray}
    where $U_n(\cos k)$ are the Chebyshev polynomials of the second kind defined in (\ref{eq:chebyshevcosk}).
    From the characteristic equation for the whole system, $W_N^1(\lambda,\cos k)=0$, one arrives at (\ref{eq:ksolutionsssh4}-\ref{eq:phissh4}) with
    \begin{equation}
    \alpha_1(\lambda)=\frac{t_m^2}{T}\chi_{1K}^1(\lambda),
    \label{eq:alphaisshm}
    \end{equation}
    where the kernel polynomial $\chi_{1K}^1(\lambda)$ is constructed by taking the first and last columns and rows in $\chi_1^1(\lambda)$ in (\ref{eq:chi11isshm}),
    \begin{equation}
    \chi_{1K}^1(\lambda)=
    \begin{vmatrix}
    (\lambda-v_{m-1})&t_{m-2}&
    \\
    t_{m-2}&(\lambda-v_{m-2})&&
    \\
    &\ddots&\ddots&\ddots
    \\
    &&&(\lambda-v_3)&t_2
    \\
    &&&t_2&(\lambda-v_2)
    \end{vmatrix}.
    \label{eq:chi1k1isshm}
    \end{equation}
    
    \item 
    Solve (\ref{eq:ksolutionsssh4}) for each $\lambda_i$ band and for all $n$ to find the $k$-solutions, with $k$ defined in the RBZ, whose respective energies are given by $\lambda_i(\cos k)$.
    The form of the eigenstates in real-space is given by
    \begin{eqnarray}
    \ket{\psi_\lambda(k)}&=&\frac{\zeta_\lambda(k)}{\sqrt{N+1}}\sum\limits_{j=1}^N\ket{u_{\lambda,j}(k)},
    \label{eq:isshmeigstate}
    \\
    \ket{u_{\lambda,j}(k)}&=&
    \begin{bmatrix}
    c_\lambda^1(k)\sin[kj-\theta_\lambda^1(k)]
    \\
    c_\lambda^2(k)\sin[kj-\theta_\lambda^2(k)]
    \\
    \vdots
    \\
    c_\lambda^{m-1}(k)\sin[kj-\theta_\lambda^{m-1}(k)]
    \\
    c_\lambda^m(k)\sin[kj]
    \end{bmatrix},
    \label{eq:isshmeigstate2}
    \\
    \vert \zeta_\lambda(k)\vert^2&=&(N+1)\Big[\sum\limits_{j=1}^N\vert u_{\lambda,j}(k)\vert^2\Big]^{-1},
    \label{eq:normfactorisshm}
    \end{eqnarray}
    where the $c_\lambda^i(\lambda)$ coefficients are obtained from the eigenstate under PBC [see an example for the SSH$_4$ model in (\ref{eq:eigstatessh4pbc})] and the $\theta_\lambda^i(k)$ phases from \eqref{eq:analyphasesisshm} (anticipating some results of the next section).
    Note that we set $\theta_\lambda^m(k)=0$, which in turn defines $\theta_\lambda^1(k)\equiv\phi_\lambda^1(k)$.
    
    \item 
    If one does not find all $mN$ states with (\ref{eq:ksolutionsssh4}) it means that there are edge states.
    These can be found by following the procedure leading to (\ref{eq:chebyshevcoshk}-\ref{eq:normobssh4}) laid out in Section~\ref{subsec:outofbandssh4}, adapting (\ref{eq:blochstateobssh4}) to the size of the chain considered.
\end{enumerate}

Note that fixing all intracell hoppings ($t_1,t_2,\dots,t_{m-1}$) and varying the intercell hopping $t_m$ in the determination of $\alpha_1(\lambda)$ in (\ref{eq:alphaisshm}) provides a practical way of crossing through different regimes in the energy spectrum, in agreement with the approach followed in Ref.[\onlinecite{Midya2018}] to detect topological transitions in some types of SSH$_4$ models.

It should also be noted that this method assumes all $t_i>0$, such that $T\neq 0$ in \eqref{eq:tprod}.
However, when one or more $t_i$ are zero, the ISSH$_m$ chain becomes simply a sequence of decoupled  and repeated small segments of few sites, whose highly degenerate eigenstates can be easily obtained.
In the specific case where at least one hopping parameter is zero \textit{but} $t_m>0$, the decoupled segments at the edge unit cells are different from those at the bulk, and may as a consequence harbor non-decaying edge states, which can be regarded as edge states with $p\to+\infty$ \cite{Kunst2017} (for instance, the fully dimerized limit of an open SSH chain in the topological phase has $t_1=0$ and $t_2>0$, leading to the appearance of zero-energy states localized at the decoupled edge sites).

A striking result of this method is that from the calculation of $\chi_{1K}^1(k)$ in (\ref{eq:chi1k1isshm}), together with calculation of the band structure under PBC, one can derive the full energy spectrum of any ISSH$_m$ model under OBC.
In a sense, $\chi_{1K}^1(k)$ codifies the relevant features of any given ISSH$_m$ model.

\section{Non-integer number of unit cells}
\label{sec:nonintegerucells}
\begin{figure*}[th]
	\begin{center}
		\includegraphics[width=0.95 \textwidth]{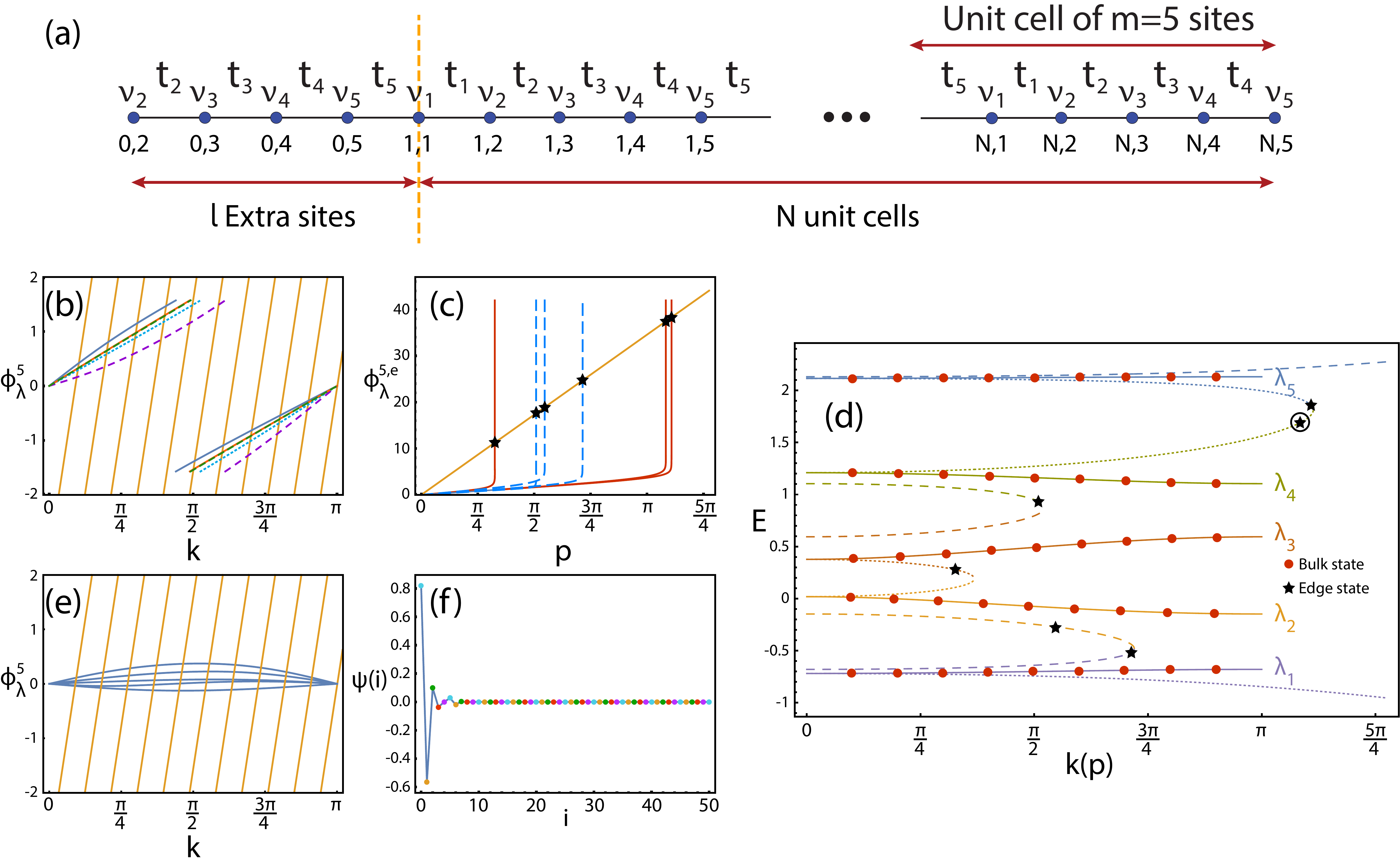}
	\end{center}
	\caption{
		(a) ISSH$_5$ chain with $N$ complete unit cells and extra sites at unit cell $j=0$. 
		(b) and (c) $\phi_\lambda^5$ and $\phi_{\lambda,\pm}^{5,e}$ phases as a function of $k$ and $p$, respectively, for the ISSH$_m$ chain in (a) with $N=10$ unit cells and an extra site at $\ket{0,5}$, $\delta=0$ and $v_i=t_i$. 
		(d) Energy spectrum of the ISSH$_5$ model with $\delta=0$ and $v_i=t_i$ in the RBZ as a function of $k$ for the $\lambda(\cos k)$ bands (solid curves) and as a function of $p$ for the $\lambda(\cosh p)$ bands (dotted curves) and $\lambda(-\cosh p)$ bands (dashed curves).
		Identically colored bands give the $\cos k$ and $\pm\cosh p$ paremetrizations of the same $\lambda$ band.
		The $k$ ($p$) solutions for an open chain with $N=10$ unit cells and an extra site at $\ket{0,5}$ are indicated by the red dots (black stars).
		Highlighted encircled state corresponds to the one depicted in (f).
	    (e) Same as in (b) but for $\delta=\frac{\pi}{2}$.
        (f) Spatial profile of the highlighted edge state in (d) along the $i$ sites of the chain, with maximum amplitude at the extra site in $i=0$, computed using \eqref{eq:eigstateobsshm} and verified numerically.}
	\label{fig:issh5}
\end{figure*}
So far we have restricted our studies to open ISSH$_m$ models with $N\in \mathbb{N}$ unit cells, implying a site 1 and a site $m$ at opposite edges, as shown for the ISSH$_5$ model in Fig.~\ref{fig:issh5}(a).
In this section we will determine the general solutions for arbitrary terminations of the ISSH$_m$ model.
We choose to fix the right edge at the $\ket{N,m}$ site, so that the last $m$ sites define the unit cell, and vary the terminations by adding $l$ sites, with $l=1,2,\dots,m-1$, in the unit cell 0 at the left edge, as exemplified for the ISSH$_5$ model in Fig.~\ref{fig:issh5}(a).
For instance, adding $l=1,2,3$ sites in the SSH$_4$ model enlarges the Hamiltonian in (\ref{eq:hamiltssh4}) at the bottom by $l$ rows and columns.
It should be noted that this exhausts all different possibilities, since adding sites also at the right edge just amounts to a redefinition of the unit cell and, therefore, of the hopping and on-site potential parameters, such that one effectively is adding sites at the left edge.

\begin{figure}[th]
	\begin{center}
		\includegraphics[width=0.47 \textwidth]{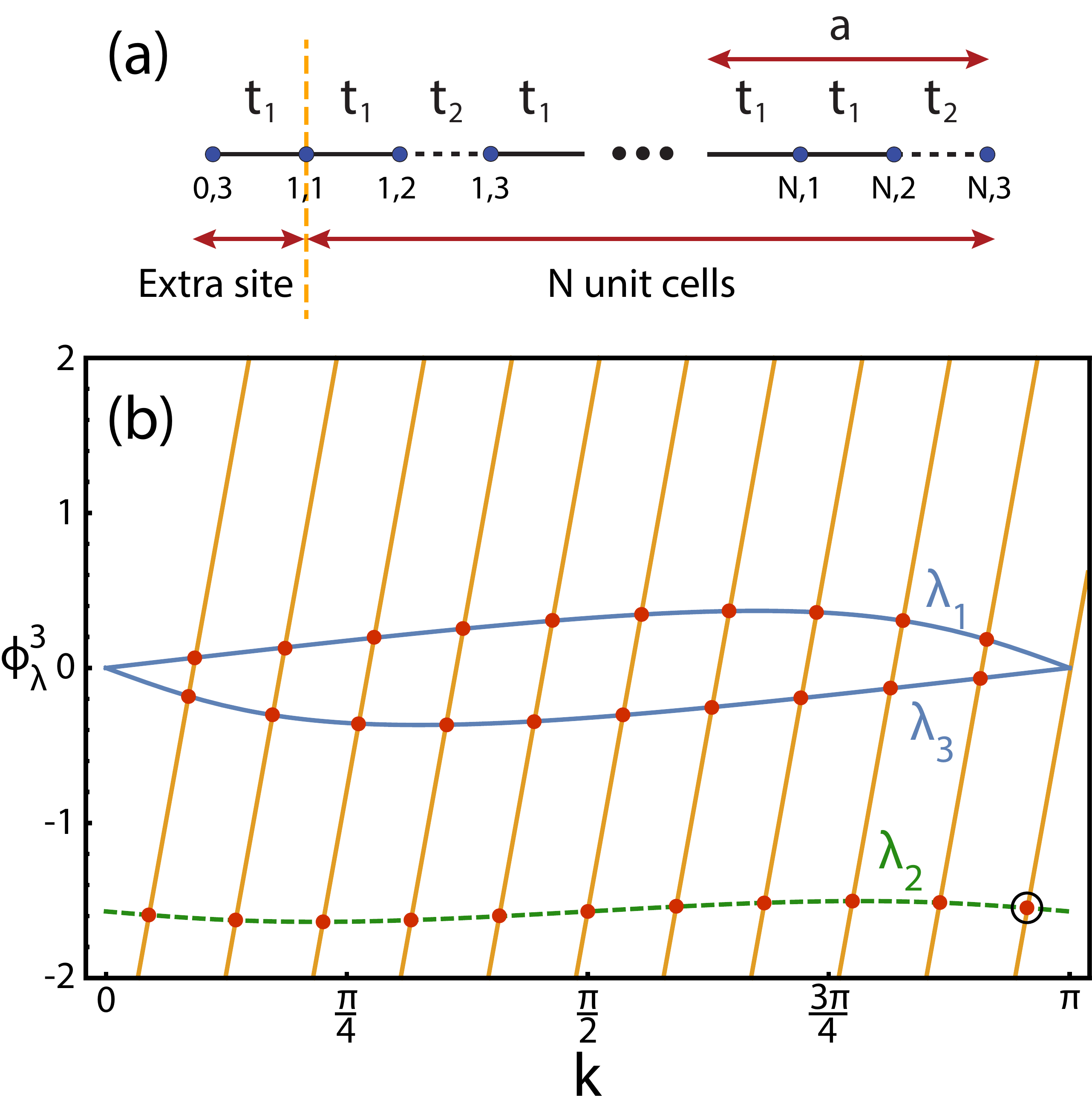}
	\end{center}
	\caption{
		(a) $t_1t_2t_1$ chain with $N$ complete unit cells and an extra site at unit cell $j=0$.
		(b) $\phi_\lambda^3$ phases as a function of $k$ for the chain in (a) with $N=10$ complete unit cells and $t_2/t_1=0.5$.
	    The highlighted encircled solution in band 2 comes from the $n=N+1$ equation for $\lambda_2$ in \eqref{eq:ksolutionsisshmi}, meaning there is one more solution for band 2 than for bands 1 and 3.}
	\label{fig:t1t2t1plus1}
\end{figure}

In general, the characteristic equation for the ISSH$_m$ model with a $\ket{0,i}$ site at the left edge is defined as $\chi_{N+1}^i=0$.
All equations in (\ref{eq:chiissh1}-\ref{eq:chiisshm}) can be developed to the form of (\ref{eq:chi1isshm}),
\begin{eqnarray}
\chi_n^i(\lambda)&=&2T\cos k\chi_{n-1}^i(\lambda)-T^2\chi_{n-2}^i(\lambda),
\label{eq:chiiisshm}
\\
W_n^i(\lambda,\cos k)&:=&T^{N-n}\chi_n^i(\lambda),
\label{eq:chiinisshm}
\\
W_n^i&=&2\cos kW_{n-1}^i-W_{n-2}^i,
\end{eqnarray}
with $i=2,\dots,m$ and $T$ defined in (\ref{eq:tprod}).
In order to express $W_n^i$ in terms of Chebyshev polynomials $U_n$, we compute $W_1^i=T^{N-1}\chi_1^i$ and $W_2^i=T^{N-2}\chi_2^i$ to find, using the same inductive reasoning followed in (\ref{eq:w0ssh4}-\ref{eq:modchebyshevssh4}),
    \begin{eqnarray}
W_n^i&=&T^{N-1}\big[\chi_1^i U_{n-1}+\frac{T_{i-1,m}^2}{T}\chi_{1K\uparrow}^{m+2-i}U_{n-2}\big]
\label{eq:modchebyshevisshmi},
\\
T_{i-1,m}^2&=&\prod\limits_{s=i-1}^mt_s^2,   
\end{eqnarray}
where $\chi_{1K\uparrow}^{i}$ is the kernel determinant of $\chi_{1\uparrow}^{i}$, which is the bottom up expansion of the characteristic polynomial, e.g., for the ISSH$_4$ model one has
\begin{eqnarray}
\chi_{1}^2(\lambda)&=&\chi_{_{1:3}}(\lambda)=
\begin{vmatrix}
\lambda-v_4&t_3&0
\\
t_3&\lambda-v_3&t_2
\\
0&t_2&\lambda-v_2
\end{vmatrix},
\\
\chi_{1\uparrow}^2(\lambda)&=&\chi_{_{4N-2:4N}}(\lambda)=
\begin{vmatrix}
\lambda-v_3&t_2&0
\\
t_2&\lambda-v_2&t_1
\\
0&t_1&\lambda-v_1
\end{vmatrix},
\end{eqnarray}
such that $\chi_{1K}^{2}=\lambda-v_3$ and $\chi_{1K\uparrow}^{2}=\lambda-v_2$.
The boundary conditions are defined as $\chi_{1K\uparrow}^{m-1}=1$ and $\chi_{1K\uparrow}^{m}=0$.
The characteristic equation can be written as
\begin{eqnarray}
W_{N+1}^i(\lambda,\cos k)&=&0, \nonumber
\\
U_n(\cos k)+\alpha_i(\lambda)U_{n-1}(\cos k)&=&0,
\label{eq:characteristiceqisshmi}
\end{eqnarray}
where
\begin{equation}
\alpha_i(\lambda)=\frac{T_{i-1,m}^2}{T}\frac{\chi_{1K\uparrow}^{m+2-i}}{\chi_1^i}.
\label{eq:alphai}
\end{equation}
It is clear that (\ref{eq:characteristiceqisshmi}) leads to the solution of (\ref{eq:cotssh4}) with $\alpha_1(\lambda)\to\alpha_i(\lambda)$, so that (\ref{eq:ksolutionsssh4}-\ref{eq:phissh4}) become
\begin{eqnarray}
\phi^i_\lambda(k)&=&(N+1)k-n\pi,\ \ \ n=1,2,\dots,N+1,
\label{eq:ksolutionsisshmi}
\\
\phi_\lambda^i(k)&=&\cot^{-1}\Big[\frac{1}{\alpha_i(\lambda)\sin k}+\cot k\Big], 
\label{eq:phiisshmi}
\end{eqnarray}
where now there is an extra equation relative to $n=N+1$. However, for a system with $l$ extra sites there is at most $l$ bands with $N+1$ bulk state solutions, such that no more than $mN+l$ states are found with (\ref{eq:ksolutionsisshmi}), as expected.
This is illustrated for the $t_1t_2t_1$ model \cite{Alvarez2019}, which is an SSH$_3$ model with $t_1=t_3$, with an extra site shown in Fig.~\ref{fig:t1t2t1plus1}(a).
In-gap topological edge states appear in this model when $\vert t_1\vert>\vert t_2\vert$ and there is at least one edge with a single $t_1$ hopping followed by $t_2$ \cite{Marques2017}.
As the extra site of Fig.~\ref{fig:t1t2t1plus1}(a) generates two consecutive $t_1$ hoppings at the left edge, we expect all $3N+1$ states to be bulk states.
Indeed, $3N+1$ $k$-solutions are found in Fig.~\ref{fig:t1t2t1plus1}(b), where it can be seen that the top ($\lambda_3$) and bottom ($\lambda_1$) energy bands yield $N$ solutions each, whereas the middle band ($\lambda_2$) yields $N+1$ solutions, with the extra one coming from the $n=N+1$ equation in (\ref{eq:ksolutionsisshmi}).

Concerning possible edge states, they can be found with the same relations (\ref{eq:psolutionsssh4}-\ref{eq:phiobssh4}) found for the $l=0$ case,  with the $\alpha^{e}_{1,\pm}(\lambda)\to\alpha^{e}_{i,\pm}(\lambda)$ and $\phi_{\lambda,\pm}^{1,e}(p)\to\phi_{\lambda,\pm}^{i,e}(p)$ substitutions.

\subsection{ISSH$_m$ with $l=1$}
\label{sec:isshml1}

Let us now turn again to the ISSH$_5$ model of Fig.~\ref{fig:issh5}(a) and study separately the $l=1$ and $l=4$ cases, which exemplify the different behaviors a general ISSH$_m$ model can manifest when extra sites are added.

When $l=1$, the chain in Fig.~\ref{fig:issh5}(a) ends with a $\ket{0,5}$ site at the left edge.
The explicit expression for $\alpha_5(\lambda)$ can be calculated from (\ref{eq:alphai}),
\begin{equation}
\alpha_5(\lambda)=\frac{t_4t_5}{t_1t_2t_3}\frac{(\lambda-v_2)(\lambda-v_3)-t_2^2}{\lambda-v_5},
\end{equation}
where $\chi_{1K\uparrow}^{2}=(\lambda-v_2)(\lambda-v_3)-t_2^2$ and $\chi_1^5=\lambda-v_5$.
After substituting $\alpha_5(\lambda)$ in (\ref{eq:phiisshmi}) to find $\phi_\lambda^5(k)$, one finds the $k$-solutions for every energy band $\lambda$ [whose expressions are found from the diagonalization of the bulk Hamiltonian $\mathcal{H}(k)$] through (\ref{eq:ksolutionsisshmi}), for the set of $t_i$ and $v_i$ parameters considered.
For $(t_1,t_2,t_3,t_4,t_5)=(0.2,0.4,0.6,0.8,1)$ (that is, $\delta=0$), $v_i=t_i$ and $N=10$ unit cells, the bulk $k$-solutions are given by the intersections at Fig.~\ref{fig:issh5}(b).
The total number of states is $mN+1=51$ and each band contributes with $N-1$ solutions (recall that $k=\pi$, the $k$ value at the rightmost intersection, is outside the RBZ), totaling $m(N-1)=45$ states.
The missing six states are the edge states found in Fig.~\ref{fig:issh5}(c).
The eigenenergies are retrieved by substitution of the $k$- and $p$-solutions into their respective $\lambda_i(\cos k)$ and $\lambda_i(\pm \cos p)$ bands.
The combined energy spectrum of both bulk and edge bands is shown in Fig.~\ref{fig:issh5}(d).
The highlighted edge state with the second highest $p$ is depicted in Fig.~\ref{fig:issh5}(f), where it can be seen to decay from the left edge, with a maximum of amplitude at the extra site.

In the determination of the bulk eigenstates for the $l\geq 1$ case, modified boundary conditions have to be considered, in relation to the $l=0$ case (integer number of unit cells). 
While the right boundary condition (RBC) is still given by $\braket{N+1,1}{\psi_\lambda(k)}=0$,  the addition of extra sites changes the left boundary condition (LBC), since a node has to be imposed farther to the left as $l$ increases, and is written as $\braket{0,i-1}{\psi_\lambda(k)}=0$, where $i=m-l+1$ is the left edge site. 
If one sets the phase of the $(i-1)^{th}$-component to zero the LBC is automatically satisfied [see \eqref{eq:isshmeigstate}], and in turn the phase of the $1^{st}$-component becomes $\phi_\lambda^i(k)$, so that the solutions obtained from \eqref{eq:ksolutionsisshmi} also satisfy the RBC.
The phases of each component within the unit cell of the eigenstates for the ISSH$_5$ model with different terminations, relative to the $l=0$ case given by \eqref{eq:isshmeigstate2}, are shown in Table \ref{tab:issh5phases}.
It should be noted that it is at the level of the bulk eigenstate that the phases are set according to each case: for instance, in the SSH$_4$ model studied above the phases are set in the bulk eigenstate of \eqref{eq:eigstatessh4pbc}, \textit{before} the anti-symmetric combination of $k$-states in \eqref{eq:eigstatessh4antisym} that leads to the eigenstate under OBC, where each component becomes a sine function dependent on its phase.
\begin{table}[h]
	\begin{center}
		\begin{tabular}{|c|c|c|c|c|c|}
			\hline
			\diagbox[innerwidth=2cm,innerleftsep=.1cm,innerrightsep=10pt]{Comp.}{\\$l$}
			&0&1&2&3&4\\ 
			\hline
			1&$\theta_\lambda^1$&$\theta_\lambda^1-\theta_\lambda^4$&$\theta_\lambda^1-\theta_\lambda^3$&$\theta_\lambda^1-\theta_\lambda^2$&0
			\\ 
			\hline
			2&$\theta_\lambda^2$&$\theta_\lambda^2-\theta_\lambda^4$&$\theta_\lambda^2-\theta_\lambda^3$&0&$\theta_\lambda^2-\theta_\lambda^1$\\
			\hline
			3&$\theta_\lambda^3$&$\theta_\lambda^3-\theta_\lambda^4$&0&$\theta_\lambda^3-\theta_\lambda^2$&$\theta_\lambda^3-\theta_\lambda^1$\\
			\hline
			4&$\theta_\lambda^4$&0&$\theta_\lambda^4-\theta_\lambda^3$&$\theta_\lambda^4-\theta_\lambda^2$&$\theta_\lambda^4-\theta_\lambda^1$\\ 
			\hline
			5&0&$-\theta_\lambda^4$&$-\theta_\lambda^3$&$-\theta_\lambda^2$&$-\theta_\lambda^1$\\ 
			\hline
		\end{tabular}  
	\end{center}
	\caption{Phases of the components (Comp.) within the unit cell of the ISSH$_5$ chain with $N$ complete unit cells and $l=5+1-i$ extra sites added at the left, relative to the $l=0$ case.}
	\label{tab:issh5phases}
\end{table}
Recalling that the phase of the $1^{st}$-component equates with $\phi_\lambda^i$, one gets a system of coupled equations from which analytical expressions for all phases can be obtained,
\begin{equation}
\begin{cases}
\phi_\lambda^1=\theta_\lambda^1
\\
\phi_\lambda^5=\theta_\lambda^1-\theta_\lambda^4
\\
\phi_\lambda^4=\theta_\lambda^1-\theta_\lambda^3
\\
\phi_\lambda^3=\theta_\lambda^1-\theta_\lambda^2
\\
\phi_\lambda^2=0
\end{cases}
\Rightarrow
\begin{cases}
\theta_\lambda^1=\phi_\lambda^1
\\
\theta_\lambda^2=\phi_\lambda^1-\phi_\lambda^3
\\
\theta_\lambda^3=\phi_\lambda^1-\phi_\lambda^4
\\
\theta_\lambda^4=\phi_\lambda^1-\phi_\lambda^5
\\
\theta_\lambda^5=0
\end{cases}.
\end{equation}
These equations can be readily generalized for any ISSH$_m$ model as
\begin{equation}
\begin{cases}
\theta_\lambda^1=\phi_\lambda^1
\\
\vdots
\\
\theta_\lambda^j=\phi_\lambda^1-\phi_\lambda^{j+1}
\\
\vdots
\\
\theta_\lambda^m=0
\end{cases},
\label{eq:analyphasesisshm}
\end{equation}
with $j=2,3,\dots,m-1$.
The set of all $\{\phi_\lambda^i\}$ is obtained from \eqref{eq:phissh4} and \eqref{eq:phiisshmi}.
If, on the one hand, the $c_\lambda^i$ coefficients of the eigenstates in \eqref{eq:isshmeigstate2} can in general be easily extracted from the bulk eigenstates under PBC, on the other hand it can be numerically challenging to extract also from them all $\theta_\lambda^i(k)$ phases, which can have rather involved expressions.
As such, the ability to find analytical expressions for the phases through \eqref{eq:analyphasesisshm} can reduce significantly the computational complexity of this method. 
The general expression for the bulk eigenstates for an ISSH$_m$ chain with a node at $\ket{0,i-1}$ is given by
\begin{eqnarray}
\ket{\psi_\lambda(k)}&=&\zeta_\lambda(k)\sqrt{\frac{2}{N+1}}\times \nonumber\\
& &\bigg[\sum\limits_{j=1}^N\ket{u_{\lambda,j}(k)}+\ket{u_{\lambda,0}(k)}_{i\to m}\bigg],
\label{eq:eigstatesshm}
\end{eqnarray}
\begin{eqnarray}
\ket{u_{\lambda,j}(k)}&=&
\begin{bmatrix}
c_\lambda^1\sin[kj-\theta_\lambda^1+\theta_\lambda^{i-1}]
\\
c_\lambda^2\sin[kj-\theta_\lambda^2+\theta_\lambda^{i-1}]
\\
\vdots
\\
c^{i-1}_\lambda\sin[kj]
\\
\vdots
\\
c_\lambda^{m-1}\sin[kj-\theta_\lambda^{m-1}+\theta_\lambda^{i-1}]
\\
c_\lambda^m\sin[kj+\theta_\lambda^{i-1}]
\end{bmatrix},
\label{eq:isshmeigstateextra}
\\
\ket{u_{\lambda,0}(k)}_{i\to m}&=&
\begin{bmatrix}
c^{i}_\lambda\sin[-\theta_\lambda^{i}+\theta_\lambda^{i-1}]
\\
\vdots
\\
c_\lambda^{m-1}\sin[-\theta_\lambda^{m-1}+\theta_\lambda^{i-1}]
\\
c_\lambda^m\sin[\theta_\lambda^{i-1}]
\end{bmatrix},
\label{eq:isshmeigstateextra0}
\end{eqnarray}
\begin{eqnarray}
\vert \zeta_\lambda(k)\vert^2&=&\frac{N+1}{2}\Big[\sum\limits_{j=1}^N\vert u_{\lambda,j}(k)\vert^2+\vert u_{\lambda,0}(k)\vert_{i\to m}^2\Big]^{-1},
\label{eq:normfactorisshmextra}
\end{eqnarray}
where $\ket{u_{\lambda,0}(k)}_{i\to m}$ accounts for the extra sites at the $j=0$ unit cell.
The eigenstates of the edge states can be found, as for the $l=0$ case, by applying $k\to q+ip$, with $q=0\vee \pi$, $c_\lambda^i(k)\to c_{\lambda,\pm}^{i,e}(p)$, and $\phi_\lambda^i(k)\to i\phi_{\lambda,\pm}^{i,e}(p)$ to the phases in \eqref{eq:analyphasesisshm}, so that $\theta_\lambda^i(k)\to i\theta_\lambda^{i,e}(p)$, resulting in
	\begin{equation}
	\begin{split}
	\ket{\psi^{e}_ {\lambda,\pm}(p)}=& \zeta^e_{\lambda,\pm}(p)\times \\ & \sum\limits_{j=1}^N(\pm)^j\bigg[\ket{u^{e}_ {\lambda,j,\pm}(p)}+\ket{u^{e}_ {\lambda,0,\pm}(p)}_{i\to m}\bigg],
\label{eq:eigstateobsshm}
	\end{split}
	\end{equation}
\begin{eqnarray}
\ket{u_{\lambda,j,\pm}^e(p)}&=&
\begin{bmatrix}
c_{\lambda,\pm}^{1,e}\sinh[pj-\theta_\lambda^{1,e}+\theta_\lambda^{i-1,e}]
\\
c_{\lambda,\pm}^{2,e}\sinh[pj-\theta_\lambda^{2,e}+\theta_\lambda^{i-1,e}]
\\
\vdots
\\
c^{i-1,e}_{\lambda,\pm}\sinh[pj]
\\
\vdots
\\
c_{\lambda,\pm}^{m-1,e}\sinh[pj-\theta_\lambda^{m+1,e}+\theta_\lambda^{i-1,e}]
\\
c_{\lambda,\pm}^{m,e}\sinh[pj+\theta_\lambda^{i-1,e}]
\end{bmatrix},
\label{eq:isshmeigstateextraob}
\\
\ket{u_{\lambda,0,\pm}^e(p)}_{i\to m}&=&
\begin{bmatrix}
c^{i,e}_{\lambda,\pm}\sinh[-\theta_\lambda^{i,e}+\theta_\lambda^{i-1,e}]
\\
\vdots
\\
c_{\lambda,\pm}^{m-1,e}\sinh[-\theta_\lambda^{m-1,e}+\theta_\lambda^{i-1,e}]
\\
c_{\lambda,\pm}^{m,e}\sinh[\theta_\lambda^{i-1,e}]
\end{bmatrix},
\label{eq:isshmeigstateextraob0}
\end{eqnarray}
\begin{eqnarray}
\vert \zeta^{e}_{\lambda,\pm}(p)\vert^2&=&\Big[\sum\limits_{j=1}^N\vert u^{e}_ {\lambda,j,\pm}\vert^2+\vert u_{\lambda,0,\pm}^e(p)\vert_{i\to m}^2\Big]^{-1}.
\label{eq:normobsshm}
\end{eqnarray}
The eigenstate in Fig.~\ref{fig:issh5}(f) has been obtained with \eqref{eq:eigstateobsshm} and verified against numerical results.

To conclude this subsection, we study also the $l=1$ case of the ISSH$_5$ model with $(t_1,t_2,t_3,t_4,t_5)=(1,0.8,0.6,0.4,0.2)$ (that is, $\delta=\frac{\pi}{2}$) and $t_i=v_i$. The bulk $k$-solutions given by the intersections at Fig.~\ref{fig:issh5}(e) show that every band contributes with $N$ solutions, so there is one extra edge solution (not shown here), yielding $5N+1$ states in total.
By comparing Fig.~\ref{fig:issh5}(e) with Fig.~\ref{fig:phasessh5}(d), which shows the bulk solutions for the same model without the extra site, one sees that the main qualitative change comes from $\lambda_5$, going from contributing with $N-1$ solutions in the latter to contributing with $N$ solutions in the former.

For intermediate cases, with $l=2,3,\dots,m-2$, the solutions are found following the same procedure as for the $l=1$ case outlined here.

\subsection{ISSH$_m$ with $l=m-1$}

When $l=m-1=4$ sites are added to the ISSH$_5$ chain with $N$ unit cells, the left edge ends with at a $\ket{0,2}$ site [see Fig.~\ref{fig:issh5}(a)]. 
Setting $i=2$ in \eqref{eq:characteristiceqisshmi} we get
\begin{eqnarray}
\chi_{1K\uparrow}^{5}&=&\alpha_2(\lambda)=0,
\\
\phi_\lambda^2(k)&=&\cot^{-1}\Big[\pm\infty\Big]=0,
\label{eq:phissh5plus1_2}
\end{eqnarray}
as shown at the last column in Table~\ref{tab:issh5phases}, yielding $k=\frac{n\pi}{N+1}$, with $n=1,2,\dots,N$, for all $\lambda$ bands, that is, we recover the same $k$.solutions as for the case of a linear chain with a single hopping parameter \cite{Kunst2019}.
The nodes of this chain, $\ket{0,1}$ and $\ket{N+1,1}$, both occur on the first component of the $\ket{u_{\lambda,j}(k)}$ eigenstates, and $\braket{0,1}{u_{\lambda,j}(k)}=0$ automatically entails (\ref{eq:phissh5plus1_2}). 
In this situation the normalization factor in (\ref{eq:normfactorisshmextra}) yields $\zeta_\lambda(k)=1$.
This can be understood by looking at the SSH$_4$ model in Fig.~\ref{fig:espectrumssh4}(a): for $l=m-1=3$ added sites, the left edge corresponds to the $\ket{0,B}$ site, and the labeled A sites are the nodes, such that the periodic chain is divided in two equal open chains with $4N+3$ sites each, hence $\zeta_\lambda(k)=1$ [see discussion below (\ref{eq:normfactorssh4})].

Returning to the ISSH$_5$ chain with $l=4$ added sites at hand, one finds trivial solutions for the $l$ edge states from $\phi_\lambda^2(k)\to i\phi_{\lambda,\pm}^{2,e}(p)=0\to p=0$.
We are unable to find the correct solutions to the missing $l$ states because the role of $\chi_1^2$ in the definition of $\alpha_2(\lambda)$ [see (\ref{eq:alphai})] gets neglected given that $\chi_{1K\uparrow}^{5}=0$ in the numerator. 
Therefore, the $l=m-1$ case requires a different approach: taking advantage of having $\chi_{1K\uparrow}^{5}=0$, one directly solves the characteristic equation $W_{N+1}^i=0$ which, from (\ref{eq:modchebyshevisshmi}), reads simply as
\begin{eqnarray}
\chi_1^2(\lambda) U_{N}(\lambda,\cos k)&=&0,
\label{eq:chi12issh5}
\\
\chi_1^2(\lambda)&=&
\begin{vmatrix}
\lambda-v_5&t_4&0&0
\\
t_4&\lambda-v_4&t_3&0
\\
0&t_3&\lambda-v_3&t_2
\\
0&0&t_2&\lambda-v_2
\end{vmatrix},
\end{eqnarray}
where $\chi_1^2(\lambda)$ is an $l^{th}$-degree polynomial. From $U_N=0$ one finds the abovementioned $k=\frac{n\pi}{N+1}$ solutions, whereas from $\chi_1^2(\lambda)=0$ the missing $l$ solutions are found, each of which can a be real (bulk) or complex (edge) $k$ (in the latter case the real part is again either 0 or $\pi$).
For the $t_1t_2t_1$ model of Fig.~\ref{fig:t1t2t1plus1}(a) with $l=2$ extra sites ($\ket{0,2}$ site at the left edge) and $t_1/t_2=2$, both solutions are bulk states and are associated with the middle $\lambda_2$ band as 
\begin{equation}
\chi_1^2(\lambda_2)=\lambda_2-t_2^2=0\to k=0,\pi.
\end{equation}

For the ISSH$_5$ chain with $l=4$, $\delta=0$ and $t_i=v_i$, all $l$ extra states are edge states, whose explicit complex $k$-values are shown in Table~\ref{tab:issh5plus4edgestates}.
\begin{table}[h]
	\begin{center}
		\begin{tabular}{|c|c|c|}
			\hline
			Band&$q$&$p$\\ 
			\hline
			$\lambda_2$&$\pi$&1.71844
			\\ 
			\hline
			$\lambda_3$&0&1.02436\\
			\hline
			$\lambda_3$&0&3.48218\\
			\hline
			$\lambda_4$&$\pi$&1.59907\\ 
			\hline
		\end{tabular}  
	\end{center}
	\caption{Edge state solutions with $k=q+ip$ for the ISSH$_5$ chain with $l=4$ extra sites, $\delta=0$ and $t_i=v_i$, obtained from $\chi_1^2(\lambda_i)=0$ in (\ref{eq:chi12issh5}), with the labeling of the $\lambda_i$ bands following Fig.~\ref{fig:issh5}(d).}
	\label{tab:issh5plus4edgestates}
\end{table}
The general form of the edge states found for $l=m-1$ is given by
\begin{eqnarray}
\ket{\psi^{e}_ {\lambda,\pm}(\sigma p)}&=&\zeta^e_{\lambda,\pm}(\sigma p) \sum\limits_{j=0}^N(\pm)^je^{-\sigma pj}\ket{u^{e}_ {\lambda,j,\pm}(\sigma p)},
\label{eq:eigstateobsshm}
\\
\ket{u_{\lambda,j,\pm}^e(\sigma p)}&=&
\begin{bmatrix}
0
\\
c_{\lambda,\pm}^{2,e}(\sigma p)e^{\sigma \theta_{\lambda,\pm}^{2,e}(p)}
\\
\vdots
\\
c_{\lambda,\pm}^{m,e}(\sigma p)e^{\sigma \theta_{\lambda,\pm}^{m,e}(p)}
\end{bmatrix},
\label{eq:isshmeigstateextramminus1ob}
\\
\vert \zeta^{e}_{\lambda,\pm}(\sigma p)\vert^2&=&\Big[\sum\limits_{j=1}^N\vert u^{e}_ {\lambda,j,\pm}\vert^2\Big]^{-1},
\label{eq:normobsshmextramminus1}
\end{eqnarray}
where $p>0$ and $\sigma=\pm$, that is, one chooses the sign $\sigma$ according to the substitution $k\to q+\sigma p$ to the bulk eigenstate [see an example for the SSH$_4$ model in \eqref{eq:eigstatessh4pbc}] $\ket{\varphi_\lambda(k)}\to\ket{u_{\lambda,j,\pm}^e(\sigma p)}$ that yields $c_{\lambda,\pm}^{1,e}(\sigma p)=0$, since the virtual sites $\ket{0,1}$ and $\ket{N+1,1}$ are both at the first component and, therefore, an edge state can only be constructed by imposing nodes at this component.
Note that $\sigma$, through its presence at the argument of the exponential is \eqref{eq:eigstateobsshm}, also defines the edge to which the state is localized: for $\sigma=(-)+$ we have a (left-) right-edge localized state.

\section{ISSH$_m$ chain connected to cluster}
\label{sec:cluster}
\begin{figure}[th]
	\begin{center}
		\includegraphics[width=0.47 \textwidth]{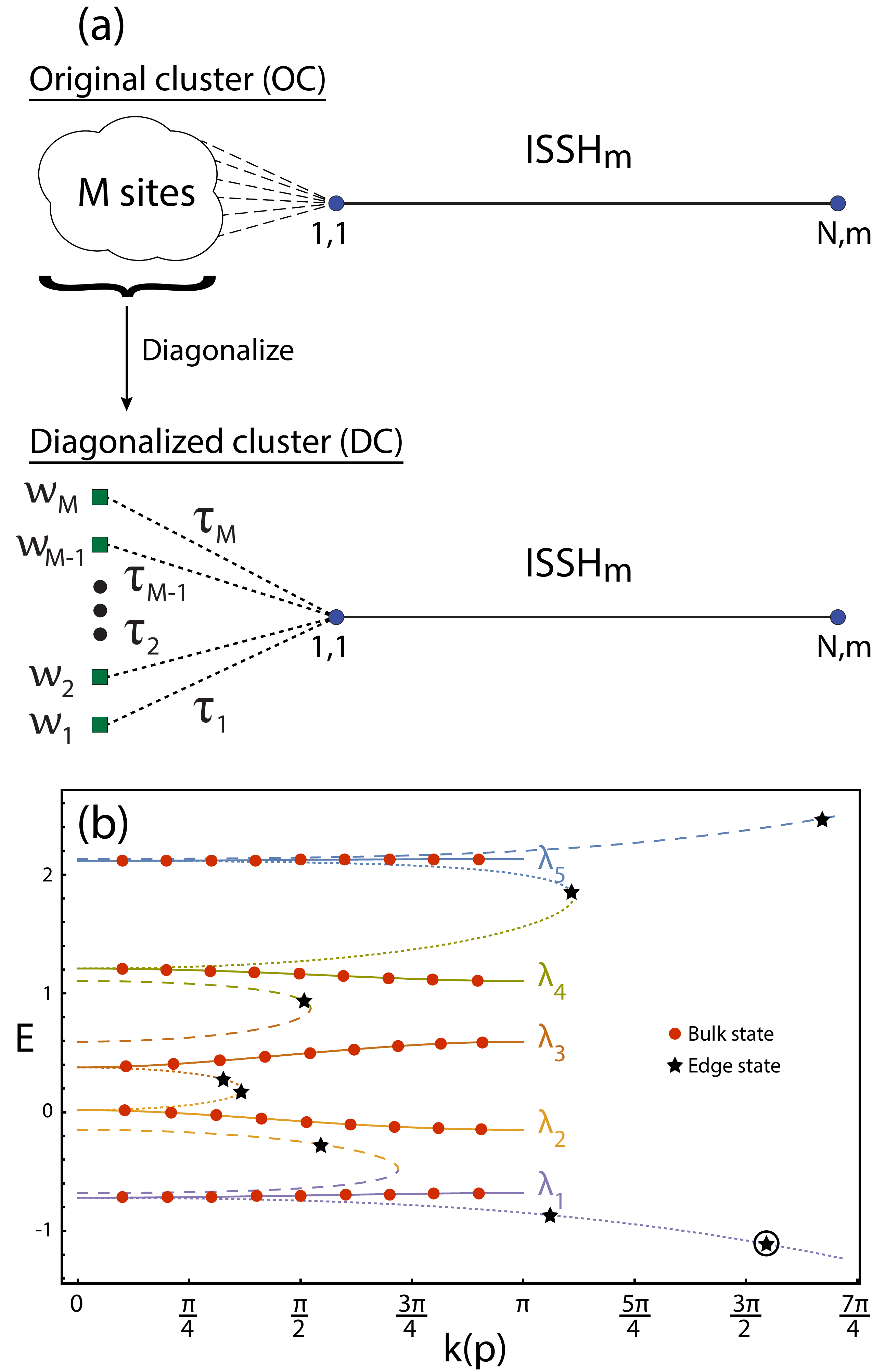}
	\end{center}
	\caption{(a) Depiction of an ISSH$_m$ chain with $N$ unit cells coupled to an arbitrary $M$-site cluster at the left edge site, before and after diagonalization of the cluster subsystem, where the diagonalized cluster states have eigenenergies $u_i$ and effective $\tau_i$ couplings to $\ket{1,1}$, with $i=1,2,\dots,M$.
		(b) Energy spectrum of the ISSH$_5$ model with $\delta=0$ and $v_i=t_i$, connected by $\ket{1,1}$ to a 3-site cluster with diagonalized parameters $(\tau_1,\tau_2,\tau_3)=(0.2,0.6,1)$ and $(w_1,w_2,w_3)=(-1,-0.5,2)$, in the RBZ as a function of $k$ for the $\lambda(\cos k)$ bands (solid curves) and as a function of $p$ for the $\lambda(\cosh p)$ bands (dotted curves) and $\lambda(-\cosh p)$ bands (dashed curves).
		Identically colored bands give the $\cos k$ and $\pm\cosh p$ paremetrizations of the same $\lambda$ band.
		The $k$ ($p$) solutions considering a chain of $N=10$ unit cells are indicated by the red dots (black stars).
		Highlighted encircled state corresponds to that of Fig.~\ref{fig:clusteredgestate}.}
	\label{fig:3sitecluster}
\end{figure}

We conclude the exposition of our method with a problem that showcases its effectiveness in dealing with a wider range of systems. Namely, we will study next a system composed of an ISSH$_m$ chain connected at one end to an $M$-site cluster with arbitrary hopping parameters and on-site potentials.
The first step in solving this problem is to independently diagonalize the $M$-site cluster, as illustrated in Fig.~\ref{fig:3sitecluster}(a). 
Then, one computes the effective couplings $\tau_j$ between these diagonalized states (which are a normalized linear combination of the original cluster sites) with energies $w_j$, where $j=1,2,\dots,M$, and the left edge site of the ISSH$_m$ chain.
The resulting characteristic polynomial reads as (except when deemed necessary, we drop the $\lambda$ and all other dependencies henceforth to ease the notation)
\begin{widetext}
	\begin{equation}
	\xi_M=\chi_{1:mN+M}=
	\begin{vmatrix}
	\lambda-v_m&t_{m-1}&&&
	\\
	t_{m-1}&\lambda-v_{m-1}&&
	\\
	&\ddots&\ddots&\ddots
	\\
	&&&\lambda-v_2&t_1
	\\
	&&&t_1&\lambda-v_1&\tau_1&\tau_2&\dots&\tau_{_{M-1}}&\tau_{_M}
	\\
	&&&&\tau_1&\lambda-w_1
	\\
	&&&&\tau_2&&\lambda-w_2
	\\
	&&&&\vdots&&&\ddots&
	\\
	&&&&\tau_{_{M-1}}&&&&\lambda-w_{_{M-1}}
	\\
	&&&&\tau_{_M}&&&&&\lambda-w_{_M}
	\end{vmatrix},
	\label{eq:xiM}
	\end{equation}
\end{widetext}
where we have defined $\xi_j\equiv\chi_{1:mN+j}$.
Expanding $\xi_j$ from below yields, after some straightforward algebra, the following recurrence relation,
\begin{equation}
\xi_j=(\lambda-w_i)\xi_{j-1}-\tau_j^2\prod\limits_{i=1}^{j-1}(\lambda-w_i)\chi_N^2,
\label{eq:xijcluster}
\end{equation}
with boundaries $\xi_0\equiv\chi_N^1$ and $\xi_{-1}\equiv\chi_N^2$.
From \eqref{eq:xijcluster}, the expression for $\xi_M$ can be found by recursively substituting the lower degree polynomials down to $j=1$,
\begin{eqnarray}
\xi_M&=&P\chi_N^1-S\chi_N^2,
\label{eq:xiMcluster}
\\
P&=&\prod\limits_{j=1}^M(\lambda-w_j),
\\
S&=&\sum\limits_{j=1}^M\tau_j^2\prod\limits_{i\neq j}(\lambda-w_i).
\label{eq:S}
\end{eqnarray}
From \eqref{eq:chi1nisshm} and \eqref{eq:chiinisshm} we have that $\chi_N^1=W_N^1$ and $\chi_N^2=W_N^2$ which, from \eqref{eq:modchebyshevisshm} and \eqref{eq:modchebyshevisshmi}, read as 
\begin{eqnarray}
\chi_N^1&=&T^{N-1}\big[TU_N-\alpha_1 U_{N-1}\big],
\label{eq:chi1nisshmcluster}
\\
\chi_N^2&=&T^{N-1}\chi_1^2U_{N-1},
\label{eq:chi2nisshmcluster}
\end{eqnarray}
Upon substituting these equations back in \eqref{eq:xiMcluster} we arrive at
\begin{equation}
\xi_M=T^{N-1}\big[PTU_N-(\alpha_1 PT-\chi_1^2S)U_{N-1}\big].
\end{equation}
Finally, with the expression for the Chebyshev polynomials given in \eqref{eq:chebyshevcosk}, the characteristic equation $\xi_M=0$ can be manipulated to yield
\begin{eqnarray}
\phi^M_\lambda&=&(N+1)k-n\pi,\ n=1,2,\dots,N+1,
\label{eq:ksolutionscluster}
\\
\phi_\lambda^M&=&\cot^{-1}\Big[\frac{1}{\alpha_M(\lambda)\sin k}+\cot k\Big], 
\\
\alpha_M(\lambda)&=&\alpha_1(\lambda)-\frac{\chi_1^2S}{PT}.
\label{eq:phicluster}
\end{eqnarray}
The expression for $\alpha_M$ shows that when the whole cluster is decoupled from the chain then all $\tau_j=0$, yielding $S=0$ and $\alpha_M=\alpha_1$ [given by \eqref{eq:alphaisshm}], that is, one is effectively finding the solutions for the decoupled ISSH$_m$ chain.
Furthermore, the signs (or more generally the phases) of the $\tau_j$ hoppings are irrelevant, as only their squared values appear in $S$.
It is also clear from \eqref{eq:S} that the labeling of the diagonalized cluster (DC) states follows an arbitrary order.
The edge states are found through (\ref{eq:psolutionsssh4}-\ref{eq:phiobssh4}) with the $\alpha^{e}_{1,\pm}(\lambda)\to\alpha^{e}_{M,\pm}(\lambda)$ substitution, where
\begin{equation}
\alpha^{e}_{M,\pm}(\lambda)=\alpha^{e}_{1,\pm}(\lambda)-\frac{\chi_1^2(\lambda)S}{PT},
\end{equation}
and  recalling that $\lambda=\lambda(\pm\cosh p)$ in this case.

If we suppose now a cluster constituted of a single site with on-site energy $v_m$ connected to $\ket{1,1}$ by $t_m$, then the problem is reduced to the $l=1$ case described in the previous section and $\phi_\lambda^M\to\phi_\lambda^m$.
The LBC in this case is given by $\braket{0,m-1}{\psi_\lambda(k)}=0$.
The same LBC holds, however, for a cluster of arbitrary size and parameters.
In a sense, all DC sites of the cluster are condensed to the $\ket{0,m}$ site, and can be thought of as an internal degree of freedom relative to this site only, whose presence modifies $\phi_\lambda^m\to\phi_\lambda^M=\phi_\lambda^m+\Delta\phi_\lambda$.
The deviation from the $l=1$ case, represented here by $\Delta\phi_\lambda=\phi_\lambda^M-\phi_\lambda^m$, propagates to every component of the eigenstate, whose bulk-periodic part can be written, after setting the phase of the $(m-1)^{th}$-component to zero (see $l=1$ case in Table~\ref{tab:issh5phases}) and using \eqref{eq:analyphasesisshm}, as
\begin{equation}
\begin{bmatrix}
c_\lambda^1e^{-i\phi_\lambda^m}
\\
c_\lambda^2e^{-i(\phi_\lambda^m-\phi_\lambda^3)}
\\
\vdots
\\
c_\lambda^{m-1}
\\
c_\lambda^me^{-i(\phi_\lambda^m-\phi_\lambda^1)}
\end{bmatrix}
\xrightarrow[]{\big(\times e^{-i\Delta\phi_\lambda}\big)}
\begin{bmatrix}
c_\lambda^1e^{-i\phi_\lambda^M}
\\
c_\lambda^2e^{-i(\phi_\lambda^M-\phi_\lambda^3)}
\\
\vdots
\\
c_\lambda^{m-1}e^{-i\Delta\phi_\lambda}
\\
c_\lambda^me^{-i(\phi_\lambda^M-\phi_\lambda^1)}
\end{bmatrix}.
\end{equation}
Following the procedure outlined in Section~\ref{sec:ssh4} of combining anti-symmetric $k$-states in order to define the eigenstates under OBC, we arrive at the following expression for the eigenstates in each $j=1,2,\dots,N$ unit cell of the ISSH$_m$ chain,
\begin{equation}
\ket{u_{\lambda,j}(k)}=
\begin{bmatrix}
c_\lambda^1\sin[kj-\phi_\lambda^M]
\\
c_\lambda^2\sin[kj-\phi_\lambda^M+\phi_\lambda^3]
\\
\vdots
\\
c^{m-1}_\lambda\sin[kj-\Delta\phi_\lambda]
\\
c_\lambda^m\sin[kj-\phi_\lambda^M+\phi_\lambda^1]
\end{bmatrix}.
\end{equation}
Since the diagonalized cluster sites are all connected to $\ket{1,1}$, whose component is given by $\psi_\lambda^{1,1}(k)=\braket{1,1}{u_{\lambda,1}(k)}=c_\lambda^1\sin[k-\phi_\lambda^M]$, the components of the eigenstates in the DC sites are directly extracted from their TB equations,
\begin{equation}
	\mu_{\lambda}^i(k)=-\frac{\tau_i}{\lambda(\cos k)-w_i}\psi_\lambda^{1,1}(k),
	\label{eq:diagclustercomponents}
\end{equation}
with $i=1,2,\dots,M$, and collected as a vector of the form $\ket{\mu_\lambda(k)}=\big(\mu_{\lambda}^1,\mu_{\lambda}^2,\dots,\mu_{\lambda}^M\big)^T$.
Finally, the full eigenstate is obtained by gathering the components relative to the ISSH$_m$ lattice and to the diagonalized cluster sites, and normalizing the resulting state, 
\begin{eqnarray}
\ket{\psi_\lambda(k)}&=&\zeta_\lambda(k)\sqrt{\frac{2}{N+1}}\bigg[\sum\limits_{j=1}^N\ket{u_{\lambda,j}(k)}+\ket{\mu_{\lambda}(k)}\bigg],
\label{eq:eigstatecluster}
\\
\vert \zeta_\lambda(k)\vert^2&=&\frac{N+1}{2}\Big[\sum\limits_{j=1}^N\vert u_{\lambda,j}(k)\vert^2+\vert \mu_{\lambda}(k)\vert^2\Big]^{-1}.
\label{eq:normfactorcluster}
\end{eqnarray}
Regarding the edge states, both those decaying from the left edge cluster and those decaying from the right edge, the procedure is the same as before, that is, one applies the substitutions $k\to q+ip$, with $q=0\vee \pi$, $c_\lambda^i(k)\to c_{\lambda,\pm}^{i,e}(p)$, and $\phi_\lambda^i(k)\to i\phi_{\lambda,\pm}^{i,e}(p)$, to arrive at
\begin{equation}
\ket{u_{\lambda,j,\pm}^e(p)}=
\begin{bmatrix}
c_{\lambda,\pm}^{1,e}\sinh[pj-\phi_{\lambda,\pm}^{M,e}]
\\
c_{\lambda,\pm}^{2,e}\sinh[pj-\phi_{\lambda,\pm}^{M,e}+\phi_{\lambda,\pm}^{3,e}]
\\
\vdots
\\
c^{m-1,e}_{\lambda,\pm}\sinh[pj-\Delta\phi_{\lambda,\pm}^{e}]
\\
c_{\lambda,\pm}^{m,e}\sinh[pj-\phi_{\lambda,\pm}^{M,e}+\phi_{\lambda,\pm}^{1,e}]
\end{bmatrix},
\end{equation}
from where we get the component of the state at the DC sites through
\begin{eqnarray}
	\psi_{\lambda,\pm}^{1,1}&=&\braket{1,1}{u_{\lambda,1,\pm}^e(k)}=c_{\lambda,\pm}^{1,e}\sinh[p-\phi_{\lambda,\pm}^{M,e}],
	\\
	\mu_{\lambda,\pm}^{i,e}(p)&=&-\frac{\tau_i}{\lambda(\pm\cosh p)-w_i}\psi_{\lambda,\pm}^{1,1}(p),
\end{eqnarray} 
all of which collected in $\ket{\mu_{\lambda,\pm}^{e}(p)}=\big(\mu_{\lambda,\pm}^{1,e},\mu_{\lambda,\pm}^{2,e},\dots,\mu_{\lambda,\pm}^{M,e}\big)^T$.
The complete normalized edge eigenstates are finally given by
\begin{eqnarray}
\ket{\psi_{\lambda,\pm}^e(p)}&=&\zeta_{\lambda,\pm}^e(p)\sqrt{\frac{2}{N+1}} \times \nonumber\\
& & \bigg[\sum\limits_{j=1}^N\ket{u_{\lambda,\pm,j}^e(p)}+\ket{\mu_{\lambda,\pm}^e(p)}\bigg],
\label{eq:edgeeigstatecluster}
\\
\vert \zeta_{\lambda,\pm}^e(p)\vert^2&=&\frac{N+1}{2}\Big[\sum\limits_{j=1}^N\vert u_{\lambda,\pm,j}^e(p)\vert^2+\vert \mu_{\lambda,\pm}^e(p)\vert^2\Big]^{-1}.
\label{eq:edgenormfactorcluster}
\end{eqnarray}
With all eigenstates determined, the last step is to revert back from the $M$ diagonalized to the $M$ original cluster sites.  
The components of the eigenstate in the original cluster (OC) sites can be found by solving a system of $M$ equations and $M$ variables which, in matrix notation, reads as
\begin{eqnarray}
	\ket{\mu_\lambda(k)}&=&\hat{R}\ket{\nu_\lambda(k)},
	\label{eq:clusterrotation}
	\\
	\hat{R}&=&
	\begin{pmatrix}
	r_1^1&r_2^1&\dots&r_{M-1}^1&r_M^1
	\\
	\vdots&&&&\vdots
	\\
	r_1^M&r_2^M&\dots&r_{M-1}^M&r_M^M
	\end{pmatrix},
	\\
	\sum\limits_{l=1}^M\vert r_l^i\vert^2&=&1, \mbox{for\ } i=1,2,\dots,M,
\end{eqnarray}
where $\ket{\mu_\lambda(k)}$ is known from \eqref{eq:diagclustercomponents}, $\ket{\nu_\lambda(k)}=\big(\nu_{\lambda}^1,\nu_{\lambda}^2,\dots,\nu_{\lambda}^M\big)^T$ is the vector form of the components at the OC sites and $\hat{R}$ the cluster diagonalization matrix. 
When inverted, \eqref{eq:clusterrotation} yields $\ket{\nu_\lambda(k)}=\hat{R}^{-1}\ket{\mu_\lambda(k)}$, such that by computing $\hat{R}^{-1}$ one finally obtains $\ket{\nu_\lambda(k)}$.
The same procedure is followed for the edge states, leading to $\ket{\nu_{\lambda,\pm}^e(p)}=\hat{R}^{-1}\ket{\mu_{\lambda,\pm}^e(p)}$.

\begin{figure}[h]
	\begin{center}
		\includegraphics[width=0.47 \textwidth]{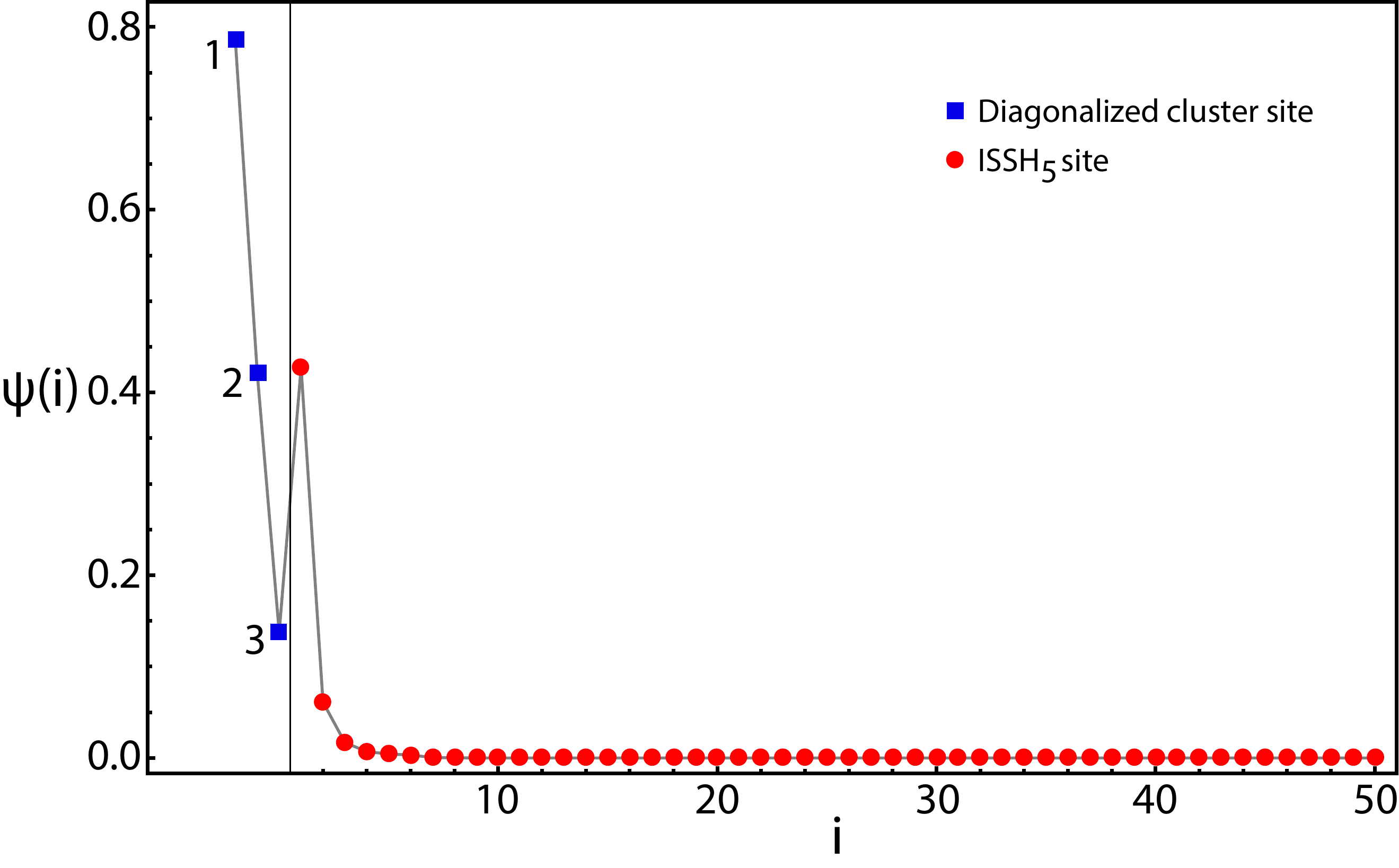}
	\end{center}
	\caption{Spatial profile of the highlighted edge state in Fig.~\ref{fig:3sitecluster}(b) along the diagonalized cluster sites (blue squares) and the $i$ sites of the ISSH$_5$ chain (red circles), computed using \eqref{eq:edgeeigstatecluster} and verified numerically.}
	\label{fig:clusteredgestate}
\end{figure}

As an example, we study a 3-site cluster of DC sites with parameters $(\tau_1,\tau_2,\tau_3)=(0.2,0.6,1)$ and $(w_1,w_2,w_3)=(-1,-0.5,2)$, connected to an ISSH$_5$ chain with $\delta=0$ and $v_i=t_i$.
Substituting the $k$- and $p$-solutions, found with \eqref{eq:ksolutionscluster} and \eqref{eq:phiobssh4}, in their respective $\lambda$ energy bands, one finds the full energy spectrum shown in Fig.~\ref{fig:3sitecluster}(b).
The edge states coinciding with those of Fig.~\ref{fig:issh5}(d) are right-edge localized, while the others are localized around the DC sites at the left-edge, such as the top and bottom edge states which have the highest $p$ values.
This is illustrated in Fig.~\ref{fig:clusteredgestate}, where the spatial profile of the lowest energy edge state in Fig.~\ref{fig:3sitecluster}(b) is shown, with the maximum of amplitude occurring at the DC site 1.

Since $p$ is the inverse localization length of the edge states, clusters with arbitrarily high $\tau_i$ and $w_i$ absolute values are expected to lead to the appearance of edge states with arbitrarily high absolute energy and $p$ values, exhibiting almost no decay to the ISSH$_m$ chain sites.
These states can only belong either to the top or to the bottom edge energy bands, which helps to explain why these bands do not have an upper limit for $p$ for any ISSH$_m$ model.
Evidently, when there are edge states with a high $p$ value, it becomes impractical to represent both the bulk and these edge states in the same energy spectrum, as has been the case with the examples studied so far.

A closer look at \eqref{eq:diagclustercomponents} shows that there are two kinds of eigenstates that cannot be found by solving  \eqref{eq:ksolutionscluster} and \eqref{eq:phiobssh4}.
The first kind is trivial: it occurs whenever a $\tau_i=0$, corresponding to a DC site decoupled from the ISSH$_m$ chain which is already an eigenstate of the overall system.
The second kind occurs when there are states with energy $\lambda=w_i$, yielding a singularity at the right-hand side of \eqref{eq:diagclustercomponents}.
These states appear when more than one DC site has energy $w_i$ and finite hoppings to $\ket{1,1}$.
The subsystem composed of these $n$-fold degenerate DC sites plus the $\ket{1,1}$ site is frustrated, that is, there are $n-1$ linear combinations of DC sites that, due to quantum interference, originate states with a node at $\ket{1,1}$ (no decay to the ISSH$_m$ chain) and degenerate energy $w_i$, somewhat akin to the ``emergent'' states studied in [\onlinecite{Alase2017}] and to the edge states with $p\to +\infty$ discussed above \cite{Kunst2017}.
The other of the $n$ states does not have a node at $\ket{1,1}$, so it appears naturally as one of the solutions of \eqref{eq:ksolutionscluster} or \eqref{eq:phiobssh4}.
For each group of $n$-fold degenerate DC sites, one finds the extra $(n-1)$-fold degenerate states by direct diagonalization of the subsystems these form with $\ket{1,1}$.

\section{Conclusions}
\label{sec:conclusions}
We developed a method for finding the anaytical solutions of ISSH$_m$ models (or commensurate Aubry-Andr\'e/Harper models) under open-boundary conditions, both for integer and non-integer number of unit cells.
It is shown that these solutions are found from self-consistent equations involving the phases of the components of the eigenstates, whose compact formulas are presented here.
The quantum number distinguishing between eigenstates in each energy band is identified with the absolute momentum defined in a reduced Brillouin zone for bulk states and a complex momentum, where the real part can only be 0 or $\pi$ and the positive imaginary part corresponds to the inverse localization length, for the edge states.
Accordingly, the concept of energy spectrum was generalized to complex momentum space in order to incorporate both bulk and edge bands simultaneously, whose visualization helps get an intuitive understanding of the system considered.
The determination of this generalized energy spectrum is not limited to the ISSH$_m$ models we study, but can be found for all open 1D models with inversion and/or time reversal symmetry, such that in the periodic model symmetric momenta are degenerate and their combination can produce nodes at specific positions, in order to satisfy open boundary conditions (as detailed in Section~\ref{sec:ssh4}).

From the ``clean'' limit, defined by unperturbed periodic modulation of the parameters across the ISSH$_m$ chain, we apply edge perturbations, in the form of arbitrary clusters connected to one of the edge sites, and find the exact analytical solutions of the whole system (ISSH$_m$ chain + cluster).
Regarding the bulk states, the role of the cluster is to induce shifts in the their absolute momentum, in relation to the ``clean'' limit.
Note that a single-site cluster essentially amounts to an edge impurity, and the exact analytical solutions derived for this case enable one to go beyond perturbation theory and consider arbitrary energy offsets for the impurity \footnote{E.g., in the studies of Ref.~[\onlinecite{Almeida2016}], a higher energy offset at the impurity of the folded chain, controlled by the central hopping term of the unfolded chain, may help enhance the upper limit of the energy gap between edge states and, consequently, of the atom-field coupling strength, while at the same time lowering the transfer time across the chain for states prepared at an edge site}.

Concerning possible applications of this method, we highlight some of them: i) as the groundwork of future studies in commensurate Aubry-Andr\'e/Harper models \cite{Lahini2009,Kraus2012,Ganeshan2013,Lang2014,Shen2014,Schreiber2015,Ke2016,Zeng2016,Cao2017,Zhao2017,Malla2018,Das2019}, ii) in the topological characterization of ISSH$_m$ models, such as the ISSH$_3$ \cite{Lang2012,Lang2014,Ke2016,Liu2017,Qin2017,Marques2017,Alvarez2019}, ISSH$_4$ \cite{Eliashvili2017,Kremer2017,Midya2018,Maffei2018,Zhang2019,Marques2019} and ISSH$_6$ \cite{Mei2012,Midya2018} models, iii) in studies on quantum state transfer across more complex ISSH$_m$ models \cite{Almeida2016,Lang2017,Mei2018,Longhi2019}, and iv) in simplifying the calculation of expectation values of arbitrary operators or interacting matrix elements in many-body problems built on these models \cite{Duncan2018}.

The results presented here lay the foundations for future studies on this topic.
We plan to extend the method to systems with \textit{both} edges of an ISSH$_m$ chain coupled to arbitrary clusters/impurities.
Although these solutions are found following the same procedure as the one outlined here, preliminary calculations show that several intermediate steps and new definitions have to be included, leading to additional terms on the characteristic equations and more complex analytical expression to the phases.
We point out that if the whole system has inversion-symmetry, then the subspaces of even and odd solutions can be decoupled from one another, each becoming an ISSH$_m$ chain connected to a cluster at a single edge, which can be solved following the steps detailed in Section~\ref{sec:cluster}.
These studies are expected to be relevant, e.g., for applications in quantum state transfer, where the dynamics across the data bus (the ISSH$_m$ chain) is controlled by external manipulations on the emitter and receiver sites (the edge impurities) \cite{Wojcik2007,Almeida2018,Junior2019} or on the edge multi-branches (clusters), allowing in this case simultaneous transfer of states \cite{Neto2013}.
Conversely, we also plan to address the case where a cluster is embedded in the middle of an ISSH$_m$ chain, in a way that preserves reflection symmetry.
This can prove useful in conductance studies on molecules or nano-rings \cite{Lopes2014,Maiti2015} coupled to \textit{finite} leads.

Another problem that we are currently addressing is the extension of the method presented here to two-dimensional (2D) lattices with a linear profile and open boundaries along both directions, such as the 2D SSH model, where a $\cos k_i$ dependence in the energy bands is preserved (required to write the characteristic equation in terms of Chebyshev polynomials), where now the momentum $\mathbf{k}=(k_1,k_2)$ is a vector.
Interesting questions arise for these higher dimensional models. 
How would the introduction of a magnetic flux through the plaquettes affect the solutions?
Can bipartite lattices with a different number of sites in each sublattice (e.g., the Lieb lattice), which entails the presence of flat bands, still be solved?
Aside form bulk and edge states, can higher-order topological (corner) states, with complex momentum in both directions, be found?
In principle, we expect a solution for these 2D lattices to be readily generalizable to models of arbitrary dimension.

\section*{Acknowledgments}
\label{sec:acknowledments}

This work is funded by FEDER funds through the COMPETE 2020 Programme
and National Funds throught FCT - Portuguese Foundation for Science
and Technology under the project UID/CTM/50025/2019 and 
under the project PTDC/FIS-MAC/29291/2017. 
AMM acknowledges financial support
from the FCT through the work contract CDL-CTTRI-147-ARH/2018, and from the Portuguese Institute for Nanostructures, Nanomodelling and Nanofabrication (I3N) through the grant BI/UI96/6376/2018. 
RGD appreciates the support by the Beijing CSRC.
AMM is grateful for useful discussions with Pedro Alves.

\bibliography{chebyshevbiblio}

\end{document}